\documentclass[12pt,preprint]{elsarticle}
\usepackage{hyperref,lineno,natbib,amsmath,amssymb,graphicx,multirow,algorithmic,color,microtype,fullpage}
\usepackage[skip=6pt]{caption}
\usepackage[title]{appendix}
\usepackage{siunitx}
\usepackage{algorithm,algorithmic}

\usepackage[resetlabels]{multibib}
\newcites{s}{References}

\newlength\myindent


\usepackage{graphicx}
\usepackage{array}

\usepackage{subfigure}
 \usepackage{siunitx}
\usepackage[utf8]{inputenc}
\usepackage[T1]{fontenc}

\usepackage{amssymb, amsmath, amsfonts}
\usepackage{outlines}

\usepackage{enumitem}
\setlist[enumerate,1]{label=\Roman*.}
\setlist[enumerate,2]{label=\roman*.}
\setlist[enumerate,3]{label=\alph*.}

\newlist{steps}{enumerate}{1}
\setlist[steps, 1]{label = Step \arabic*}

\newlist{stages}{enumerate}{1}
\setlist[stages, 1]{label = Stage \arabic*}

\journal{}

\usepackage{xcolor}
\usepackage{listings}

\renewcommand{\arraystretch}{1.5}

\colorlet{mygray}{black!30}
\colorlet{mygreen}{green!60!blue}
\colorlet{mymauve}{red!60!blue}

\definecolor{webgreen}{rgb}{0,.35,0}
\definecolor{webbrown}{rgb}{.6,0,0}
\definecolor{RoyalBlue}{rgb}{0,0,0.9}
\definecolor{mywhite}{rgb}{1.0,1.0,1.0}

\definecolor{purp}{rgb}{0.4,0.2,0.8}

\usepackage{hyperref}
\definecolor{webgreen}{rgb}{0,.35,0}
\definecolor{webbrown}{rgb}{.6,0,0}
\definecolor{RoyalBlue}{rgb}{0,0,0.9}

\newcommand{\p}{\partial}
\renewcommand{\vec}[1]{\mathbf{#1}}
\newcommand{\vx}{\vec{x}}

\newcommand{\vv}{\vec{v}}

\definecolor{dblue}{rgb}{0.1,0.3,0.65}

\definecolor{colrev}{rgb}{0.9,0,0}
\definecolor{richardson}{rgb}{0.5,0.5,0.5}

\hypersetup{
   colorlinks=true, linktocpage=true, pdfstartpage=3, pdfstartview=FitV,
   breaklinks=true, pdfpagemode=UseNone, pageanchor=true, pdfpagemode=UseOutlines,
   plainpages=false, bookmarksnumbered, bookmarksopen=true, bookmarksopenlevel=1,
   hypertexnames=true, pdfhighlight=/O,
   urlcolor=webbrown, linkcolor=RoyalBlue, citecolor=webgreen,
   pdfauthor={Jiayin Lu and Chris H. Rycroft},
   pdfsubject={Numerical methods and improvements for simulating quasi-static elastoplastic materials},
   pdfkeywords={elastoplasticity, Chorin-type projection method, finite element method, second-order projection method, multi-stage predictor-corrector scheme, adaptive Runge--Kutta method, adaptive timestepping, error control},
   pdfcreator={pdfLaTeX},
   pdfproducer={LaTeX with hyperref}
}

\newcommand{\bA}{\mathbf{A}}
\newcommand{\bC}{\mathbf{C}}
\newcommand{\bD}{\mathbf{D}}
\newcommand{\bS}{\mathbf{S}}
\newcommand{\bsig}{\boldsymbol\sigma}
\newcommand{\bome}{\boldsymbol\omega}
\newcommand{\beps}{\boldsymbol\epsilon}
\newcommand{\bPhi}{\boldsymbol\Phi}
\newcommand{\bmcL}{\boldsymbol{\mathcal{L}}}
\newcommand{\bmcH}{\boldsymbol{\mathcal{H}}}
\newcommand{\bmcS}{\boldsymbol{\mathcal{S}}}
\newcommand{\bmcV}{\boldsymbol{\mathcal{V}}}
\newcommand{\bldone}{\mathbf{1}}
\newcommand{\trans}{\mathsf{T}}

\renewcommand{\vec}[1]{\mathbf{#1}}
\newcommand{\vb}{\vec{b}}
\newcommand{\vn}{\vec{n}}
\newcommand{\vq}{\vec{q}}
\newcommand{\vw}{\vec{w}}
\newcommand{\vy}{\vec{y}}
\newcommand{\vpsi}{\boldsymbol\psi}
\newcommand{\vLambda}{\boldsymbol\Lambda}

\usepackage{amssymb}
\usepackage{bm}
\usepackage{amsmath}

\DeclareMathOperator{\Tr}{tr}

\journal{Journal of Computational Physics}

\begin{document}

\begin{frontmatter}

\title{Numerical methods and improvements for simulating quasi-static elastoplastic materials}

\author[UCLA,UW]{Jiayin Lu}
\author[UW,LBL]{Chris H. Rycroft}

\affiliation[UCLA]{organization={Department of Mathematics, University of California, Los Angeles},
  city={Los Angeles},
  state={CA},
  postcode={90095},
country={USA}}

\affiliation[UW]{organization={Department of Mathematics, University of Wisconsin--Madison},
            city={Madison},
            state={WI},
            postcode={53706},
            country={USA}}

\affiliation[LBL]{organization={Mathematics Group, Lawrence Berkeley Laboratory},
            city={Berkeley},
            state={CA},
            postcode={94720},
            country={United States}}

\begin{abstract}
  Hypo-elastoplasticity is a framework suitable for modeling the mechanics of many hard materials that have small elastic deformation and large plastic deformation. In most laboratory tests for these materials the Cauchy stress is in quasi-static equilibrium. Rycroft et al.\@ discovered a mathematical correspondence between this physical system and the incompressible Navier--Stokes equations, and developed a projection method similar to Chorin's projection method (1968) for incompressible Newtonian fluids. Here, we improve the original projection method to simulate quasi-static hypo-elastoplasticity, by making three improvements. First, drawing inspiration from the second-order projection method for incompressible Newtonian fluids, we formulate a second-order in time numerical scheme for quasi-static hypo-elastoplasticity. Second, we implement a finite element method for solving the elliptic equations in the projection step, which provides both numerical benefits and flexibility. Third, we develop an adaptive global time-stepping scheme, which can compute accurate solutions in fewer timesteps. Our numerical tests use an example physical model of a bulk metallic glass based on the shear transformation zone theory, but the numerical methods can be applied to any elastoplastic material.
\end{abstract}


\end{frontmatter}

\section{Introduction}
\label{sec:intro}

\subsection{Motivation}
Bulk metallic glasses (BMGs) are a type of amorphous solid. They are produced by rapidly cooling metallic liquids, resulting in metallic alloys with an amorphous microstructure~\cite{Luzgin2021BMG_overview}, where the atomic arrangements are random. In comparison to traditional polycrystalline metals and alloys, BMGs exhibit extreme strength and hardness, exceptional elasticity, and excellent resistance to corrosion and wear~\cite{chen2011BMG_overview}. These remarkable properties stem from their amorphous structure, which lacks grain boundaries and crystal defects that typically contribute to intergranular corrosion and weaken material strength. Additionally, recent advancements in thermoplastic forming techniques enable the economical and energy-efficient fabrication of BMGs, requiring low processing temperatures and forming pressures~\cite{Schroers_BMG_2010}. Consequently, BMGs hold considerable technological potential for many practical applications where strength and corrosion resistance are critical~\cite{WANG200445}. For instance, they find use in automobile components~\cite{lucaci2015} and structural materials~\cite{Inoue2007_BMG_applications,Kruzic2016}. They also show promise as next-generation biomaterials~\cite{Morrison2005BMG_Biomaterials}, such as in the realm of medical implants~\cite{OAK2009322,IDA2018505,SUN2022253}.

While BMGs hold significant promise for a wide range of practical applications, during loading, they can undergo highly heterogeneous plastic deformation localized in narrow-banded regions in the material, known as shear bands~\cite{HUFNAGEL20001071}. The rapid and uninhibited propagation of these shear bands can lead to sudden material fracture and catastrophic failure~\cite{SUN2015211}. However, the plastic deformation and fracture mechanics of BMGs remain poorly understood, due to their random atomic arrangements that lack conventional carriers of plastic dislocations~\cite{TREXLER2010759}. Therefore, it is essential to investigate and model the deformation behaviors of BMGs, in order to better predict material failure and design safer products. 

The mechanical deformation of BMGs is elastoplastic~\cite{Vaidyanathan2001BMG}. During loading, when the stress is below the material yield stress, elastoplastic materials deform purely elastically, meaning that they will return to their initial undeformed state upon unloading. When the yield stress is reached, elastoplastic materials start to deform plastically, where the deformation is permanent and irreversible upon unloading~\cite{Lubliner2008,Gurtin2010}. Many other materials of engineering importance have elastoplastic behavior, such as metals~\cite{Lipinski1989metal, Champion2003metal} and granular materials~\cite{Henann2013granular}. 

There are several mathematical formulations for modeling elastoplasticity~\cite{Xiao2006ElastoplasticityBS}. Each formulation defines how the elastic and plastic components of deformation interact at the microscopic scale. One formulation is the hypo-elastoplasticity framework, which assumes that the Eulerian rate of deformation tensor, $\bD$, can be decomposed additively into elastic and plastic parts, $\bD=\bD^{\text{el}}+\bD^{\text{pl}}$~\cite{truesdell1955,hill1958,prager1960}.  The framework introduces elasticity in rate form, which has known limitations, such as the dissipation or creation of energy in closed deformation cycles. For instance, residual stresses may remain after an elastic strain cycle~\cite{kojic1987studies}. Although these effects are problematic when elastic deformations become large, they are negligible for small elastic deformations~\cite{Xiao2006ElastoplasticityBS}. 
Thus, the framework is well-suited for modeling the mechanics of many hard materials that experience large plastic deformations and only small elastic deformations, like metals and BMGs. Additionally, it offers several numerical advantages: by using the Eulerian rate of deformation tensor, it allows for a fixed-grid framework, which simplifies topology and enhances ease of programming and parallelization. Wilkins was one of the first to numerically implement hypo-elastoplasticity~\cite{Wilkins1963}, and used it to model a copper plate being deformed due to an explosive charge. The framework has been employed frequently since then~\cite{udaykumar03,tran04,barton10}. Since this paper focuses solely on the hypo-elastoplasticity framework, we drop the ``hypo-'' prefix in the following sections.

Substituting the additive decomposition of $\bD$ from the elastoplasticity framework into the linear elasticity constitutive equation, along with a continuum version of Newton's second law, one can obtain a closed hyperbolic system of partial differential equations (PDEs) that can be solved explicitly via a standard finite-difference or finite-volume method~\cite{leveque_fd,leveque_fv}. The scheme needs to satisfy the Courant--Friedrichs--Lewy (CFL) condition~\cite{CFLcondition1967} for numerical stability, which requires the timestep size $\Delta t$ and grid spacing $h$ to satisfy $\Delta t\leq \frac{h}{c_e}$, where $c_e$ is a typical elastic wave speed of the material. For many materials of interest, such as metals or BMGs, $c_e$ is in the order of kilometers per second. This means that in many lab tests, the loading timescale is orders of magnitude longer than the time for elastic waves to travel through the material~\cite{Yang2005,Bouchbinder2008stabilitycavity}. Therefore, the CFL condition poses a prohibitive constraint on the explicit method to simulate realistic timescales of minutes, hours or days.

It is therefore necessary to have alternative simulation approaches that avoid resolving the elastic waves. In practical scenarios, we are often interested in the material mechanics in a much longer timescale than the elastic wave travel time. This places us in the quasi-static (QS) regime, where we look at the limit of long times and small velocities. Through a limiting procedure~\cite{Rycroft2015BMG2D}, one can show that in this regime, Newton's second law can be replaced by a divergence-free constraint on the stress for the stress to remain in QS equilibrium. However, the system is no longer a hyperbolic PDE system, and the ability to explicitly update the velocity field in time is lost.

In a relatively recent work~\cite{Rycroft2015BMG2D}, Rycroft et al.\@ discovered a mathematical correspondence of the QS elastoplastic equations to the incompressible Navier--Stokes equations~\cite{Batchelor1967fluid,acheson1990fluid}, implying that numerical methods used to solve the latter system can be transferred to solve the former system. In particular, Rycroft et al.\@ looked at Chorin's projection method~\cite{Chorin1967,Chorin1968}, a well-known numerical technique for incompressible Newtonian fluids. In the first step of the method, one explicitly updates fluid velocity in time. After this comes a projection step, where one solves an elliptic problem for the pressure term, which is used to project the velocity field to maintain the incompressibility constraint. This projection is orthogonal with respect to a suitably defined inner product. Based on the observation of a direct methematical correspondence between the stress and velocity of a QS elastoplastic solid and the velocity and pressure of the incompressible fluid, Rycroft et al.~\cite{Rycroft2015BMG2D} developed an analogous projection method for QS elastoplasticity. The method first explicitly updates the stress in time. It then proceeds to the projection step, where one solves an elliptic problem for the velocity term, which orthogonally projects the stress field to maintain the QS constraint. The connections between numerical methods for fluids and solids have been explored in different contexts before, where the fractional step approach is translated from fluid to solid, such as in a Lagrangian formulation of solid dynamics problems, specifically to handle incompressibility~\cite{sudeep_k__lahiri__2005}, for bending-dominated nearly/fully incompressible materials~\cite{GIL2014659, Haider2017, GIL2016146}. However, to the best of our knowledge, no prior works have explicitly investigated the central idea of a direct mathematical correspondence between the Navier--Stokes equations and QS elastoplasticity equations before Rycroft et al.\@~\cite{Rycroft2015BMG2D}. The mathematical connections between the two systems are reviewed in Sec.~\ref{sec:math_background}, and their respective projection methods are reviewed in Sec.~\ref{sec: numerical background}.

This above-mentioned projection method for QS elastoplasticity is well-suited to model the mechanical deformation of BMGs. BMGs typically have small elastic deformation that can be described by linear elasticity theory, but they can undergo large plastic deformation~\cite{krusiz2016BMG}, making the elastoplastic description suitable. Furthermore, their elastic moduli are on the order of \SIrange{10}{100}{GPa}, putting them in the QS physical regime in many experimental tests~\cite{Hufnagel2002BMG}. The method has been used to simulate and explain the large differences in notch fracture toughness of BMGs~\cite{rycroft12b}. It has also been used to make broad predictions about BMG toughness across a range of experimental parameters~\cite{vasoya16}, and the main predictions were subsequently verified experimentally~\cite{ketkaew18}. Boffi and Rycroft further extended the projection method to three dimensions~\cite{Boffi2020BMG3D}, and developed a coordinate transform methodology to implement Lees--Edwards boundary conditions~\cite{boffi20}, allowing for precise comparison of continuum simulations to molecular dynamics simulations~\cite{kontolati21}.

\subsection{Contribution of this paper}
In previous papers using the projection method for QS elastoplasticity, the numerical approach has been analogous to Chorin's original fluid projection method~\cite{Chorin1967,Chorin1968}. The papers employ second-order finite-difference discretization schemes in space, combined with first-order timestepping and additional first-order temporal errors from the projection. Since Chorin's original work in the late 1960's, a wide range of numerical improvements for the fluid projection method have been developed~\cite{BROWN2001projection}. The goal of this paper is to show that several of these improvements can be translated to the projection method for QS elastoplasticity, improving its performance and accuracy.

In Sec.~\ref{sec: second-order formulation} we drew inspiration from the second-order improvements of the fluid projection method. We developed a projection method with incremental velocity to solve QS elastoplasticity, which achieves second-order temporal accuracy in all solution fields. For the fluid projection method, Almgren et al.~\cite{almgren96} developed the approximate projection based on the finite element method (FEM). In Sec.~\ref{sec: FEM projection}, we implemented a finite element formulation~\cite{donea_fluid_fem,claes_fem} for the projection step of QS elastoplasticity. The resulting matrix $\mathbf{A}$ is symmetric positive-definite, and the associated linear system can be efficiently solved via different numerical techniques. The FEM also provides flexibility and robustness in setting up boundary conditions for different simulations. Lastly, in Sec.~\ref{sec:Adaptive global timestepping}, we developed an efficient adaptive timestepping scheme for the projection method, by bounding the projection size in each timestep and forcing the system to conform closely to quasi-staticity.

In Sec.~\ref{sec: numerical tests}, we numerically test the improved methods. Our numerical tests are done on a physical model of a BMG, using an athermal formulation of the shear transformation zone (STZ) theory~\cite{Falk1998,Langer2008STZ, Manning2007STZ} as a plasticity model. The STZ theory was originally guided by observations of molecular dynamics simulations~\cite{Falk2011Chi}, and has undergone substantial development~\cite{Bouchbinder2007STZ,Bouchbinder2009}. It is suitable to describe the plastic behavior for a wide variety of amorphous materials.

While we focus on simulating the mechanical deformation of BMGs in this paper, the numerical methods and improvements developed are general and can be applied to any elastoplastic materials.

\section{Mathematical preliminaries and simulation overview}
\label{sec: set up of numerical tests}
In this paper, the numerical tests all use a two-dimensional simple shear, plane strain~\cite{Barber2004planestrain} configuration. The velocity is $\vv=(u,v,0)$, and the stress tensor is
\begin{equation}
\label{eqn:stress tensor component}
\bsig=
\begin{pmatrix}
-p+s-q & \tau & 0\\
\tau & -p-s-q & 0\\
0 & 0 & -p+2q
\end{pmatrix},
\end{equation}
where $p$ is pressure, $s$ and $\tau$ are components of deviatoric stress in the $xy$ plane, and $q$ is the component of deviatoric stress in the $z$ direction out of the plane. The deviatoric stress tensor is $\bsig_0=\bsig-\frac{1}{3}\bldone(\Tr\bsig)$, and its magnitude is $\Bar{s}=|\bsig_0|=\sqrt{s^2+\tau^2+3q^2}$. Another variable of interest is the effective temperature, $\chi$, which comes from the STZ plasticity model described in detail in Sec.~\ref{sec: plasticity model and its numerical challenges}. The field variable $\chi$ describes the disorderliness of the system at the atomic scale, and increases when plastic deformation increases. Furthermore, we use a physical model of a specific BMG, Vitreloy 1~\cite{SUBHASH200233,LU20033429}. Its elasticity parameters are provided in Table~\ref{tbl:material parameters}.

\begin{table}[h!]
\centering
\begin{tabular}{|p{80mm}|p{30mm}|}
\hline
Young's modulus $E$&\SI{101}{GPa}
\\ \hline
    Poisson ratio $\nu$& 0.35
\\ \hline
    Bulk modulus $K$& \SI{122}{GPa}
\\ \hline
    Shear modulus $\mu$& \SI{37.4}{GPa}
\\ \hline
    Lam\'e's first parameter $\lambda=K-\frac{2\mu}{3}$ & \SI{97.07}{GPa}
\\ \hline
    Density $\rho_0$& \SI{6125}{kg.m^{-3}}
\\ \hline
    Shear wave speed $c_S=\sqrt{\mu/\rho_0}$& \SI{2.47}{km.s^{-1}}
\\ \hline
    Yield stress $s_Y$& \SI{0.85}{GPa}
\\ \hline

\end{tabular}
\caption{Elasticity parameters of the BMG Vitreloy 1. \label{tbl:material parameters}}
\end{table}

\subsection{Spatial discretization and finite difference stencil}
\label{sec: spatial grid and discretization}

Suppose our simulation domain is $[a_x,b_x]\times[a_y,b_y]$, which is divided into a rectangular $M \times N$ grid. Each grid cell $(i,j)$ is a square with length $\Delta x= \Delta y = h$. Field variables $\bsig$ and $\chi$ are stored at cell centers, and indexed with half-integers. The field variable $\vv$ is stored at cell corners, and indexed with full integers. Velocity boundary conditions are applied on the top and bottom walls of the simulation domain for simple shear, $\vv^B=(U,0)$ and $\vv^B=(-U,0)$, respectively. Periodic boundary conditions are imposed in the $x$ direction, and periodic images are used so that, \textit{e.g.}, a field value $f_{i,j}$ is treated as equivalent to $f_{i-M,j}$.

The spatial derivatives are approximated with a second-order finite difference method. Here, we present the relevant spatial discretization used throughout this paper. For an arbitrary field $f$, its staggered first-order derivative in the $x$ direction uses centered differencing,
\begin{equation}
\label{eqn:staggered first derivative}
\left[ \frac{\p f}{\p x} \right]_{i+\frac{1}{2},j+\frac{1}{2}}=\frac{f_{i+1,j}+f_{i+1,j+1}-f_{i,j}-f_{i,j+1}}{2h}.
\end{equation}
The second-order derivative in the $x$ direction is given by
\begin{equation}
\label{eqn:corner second derivative}
\left[ \frac{\p^2 f}{\p x^2} \right]_{i,j}=\frac{f_{i+1,j}-2f_{i,j}+f_{i-1,j}}{h^2}.
\end{equation}
The advective derivatives are upwinded for stability. Using the second-order ENO numerical scheme~\cite{Shu1988ENO}, they are given by
\begin{equation}
\label{eqn:corner advective derivative}
\left[ \frac{\p f}{\p x} \right]_{i,j}=\frac{1}{2h}
\begin{cases}
      -f_{i+2,j}+4f_{i+1,j}-3f_{i,j} & \text{if $u_{i,j}<0$  and $|[f_{xx}]_{i,j}|>|[f_{xx}]_{i+1,j}|$,}\\
      3f_{i,j}-4f_{i-1,j}+f_{i-2,j} & \text{if $u_{i,j}>0$ and $|[f_{xx}]_{i,j}|>|[f_{xx}]_{i-1,j}|$,}\\
      f_{i+1,j}-f_{i-1,j} & \text{otherwise,}
\end{cases}
\end{equation}
where $[f_{xx}]_{i,j}$ is the second-order centered difference at $(i,j)$
evaluated by Eq.~\eqref{eqn:corner second derivative}. Analogous formulas to
Eqs.~\eqref{eqn:staggered first derivative}, \eqref{eqn:corner second
derivative}, \& \eqref{eqn:corner advective derivative} are used in the $y$
direction.

\section{Mathematical background}
\label{sec:math_background}
Here, we review the mathematical connections between QS elastoplasticity and the incompressible Navier--Stokes equations.

\subsection{Elastoplasticity}
Consider an elastoplastic material with velocity
$\mathbf{v}(\mathbf{x},t)$ and Cauchy stress tensor
$\boldsymbol \sigma (\mathbf{x},t)$. The spin is
$\boldsymbol \omega =
(\nabla \mathbf{v}-(\nabla \mathbf{v})^{\mathsf{T}})/2$, and the
rate-of-deformation tensor is
$\mathbf{D}=
(\nabla \mathbf{v}+(\nabla \mathbf{v})^{\mathsf{T}})/2$. The advective
derivative for a field $f(\mathbf{x},t)$ is defined as
$\frac{df}{dt}=\frac{\partial f}{\partial t}+(\mathbf{v}\cdot \nabla )f$.
Under the assumption of small elastic deformation, the Jaumann objective
stress rate,
\begin{equation}
  \frac{\mathcal{D}\bsig}{\mathcal{D}t}=\frac{d\bsig}{dt}+\bsig\cdot\bome-\bome\cdot\bsig,
\end{equation}
describes the time-evolution of stress taking into account material translation and rotation. Note that the use of the Jaumann objective stress rate prevents an equivalent conservation law formulation, which limits this framework to smooth, continuous fields and excludes discontinuous solutions. However, as our focus is on modeling continuous deformation of the materials, this limitation does not impact the scope of our study.

Hypo-elastoplasticity assumes an Eulerian rate of deformation tensor decomposed additively into elastic and plastic parts, $\bD=\bD^{\text{el}}+\bD^{\text{pl}}$. Under this assumption, the linear elastic constitutive equation is
\begin{equation}
\label{eqn:linear constitutive eqn}
\frac{\mathcal{D}\bsig}{\mathcal{D}t}=\bC:\bD^{\text{el}}=\bC:(\bD-\bD^{\text{pl}}),
\end{equation}
where $\mathbf{C}$ is the fourth order stiffness tensor. Assuming the material
to be homogeneous and isotropic, the components of $\mathbf{C}$ are given
by
$C_{ijkl}=\lambda \delta _{ij}\delta _{kl}+\mu (\delta _{ik}\delta _{jl}+
\delta _{il}\delta _{jk})$, where $\lambda =K-(2\mu /3)$ is Lam\'e's
first parameter, $K$ is the bulk modulus and $\mu $ is the shear modulus~\cite{Lubliner2008}.
Furthermore, from Newton's second law, the velocity satisfies
\begin{equation}
\label{eqn:velocity evolution}
\rho \frac{d\vv}{dt}=\nabla \cdot \bsig,
\end{equation}
where $\rho$ is the material density. Equations \eqref{eqn:linear constitutive eqn} \& \eqref{eqn:velocity evolution} form a hyperbolic PDE system for the stress and velocity fields. They can be solved explicitly via a standard finite-difference method, but as described in the introduction, for realistic parameters, the CFL condition places severe limits on the timestep that can be used.

\subsection{QS elastoplasticity}
Now we consider deformation of the material in realistic timescales much longer than the time it takes for elastic waves to propagate through the material. Rycroft et al.~\cite{Rycroft2015BMG2D} presented the limiting procedure on scaling Eqs.~\eqref{eqn:linear constitutive eqn} \& \eqref{eqn:velocity evolution} in the limit of long times with corresponding small velocity gradients, and showed that Eq.~\eqref{eqn:velocity evolution} becomes
\begin{equation}
\label{eqn:divergence free constraint}
\nabla \cdot \bsig=\vec{0},
\end{equation}
which states that stresses are in QS equilibrium. Therefore, for the description of QS elastoplasticity, we have the constitutive equation given in Eq.~\eqref{eqn:linear constitutive eqn}, subject to the divergence-free constraint on stress given in Eq.~\eqref{eqn:divergence free constraint}. However, it is unclear how to solve the PDE system, as stress evolution in Eq.~\eqref{eqn:linear constitutive eqn} depends on velocity $\vv$ through $\bD$. The evolution equation for $\vv$ is exchanged for the constraint in Eq.~\eqref{eqn:divergence free constraint}, making it unclear how to evolve $\vv$.

\subsection{Mathematical connections to incompressible fluid dynamics}
Rycroft et al.~\cite{Rycroft2015BMG2D} discovered a surprising mathematical analogy between the QS elastoplasticity equations and the incompressible Navier--Stokes equations~\cite{Batchelor1967fluid,acheson1990fluid} for fluid dynamics. For fluid velocity $\vv$, pressure $p$, density $\rho$, and deviatoric stress $\boldsymbol\tau$ arising from viscosity, the general Navier--Stokes equations~\cite{Batchelor1967fluid,acheson1990fluid} are
\begin{equation}
\label{eqn: NS: v}
\rho \frac{d\vv}{dt}=-\nabla p+ \nabla \cdot \boldsymbol\tau,
\end{equation}
and\begin{equation}
\label{eqn: NS: rho}
\frac{d\rho}{dt} + \nabla \cdot (\rho \vv) =0.
\end{equation}
In addition, an equation of state linking $\rho$ and $p$ must be satisfied, with a typical relation being the weakly compressible model, $\rho - \rho_0 = (p - p_0)/c^2$, where $\rho_0$ is a typical density, $p_0$ is a typical pressure, and $c$ is a typical sound speed which may be very large (\textit{i.e.}\@ several kilometers per second). Similar to the elastoplasticity equations, Eqs.~\eqref{eqn: NS: v} \& \eqref{eqn: NS: rho} form a hyperbolic PDE system that can be solved explicitly with standard finite difference scheme. However, the large value of $c$ means that the CFL condition creates an extreme restriction on the numerical timestep size.

Consider the long-time limit, at a time scale much longer than the time it takes for the sound waves to propagate across the system. Following a procedure similar to elastoplastic materials in the QS limit, Eq.~\eqref{eqn: NS: rho} is exchanged for a divergence-free constraint on $\vv$,
\begin{equation}
\label{eqn:incompressible NS constraint}
\nabla \cdot \vv=0.
\end{equation}
Equations \eqref{eqn: NS: v} \& \eqref{eqn:incompressible NS constraint} are the incompressible Navier--Stokes equations. Similar to the QS elastoplasticity equations, it is unclear on how to solve them. The evolution of $\vv$ in Eq.~\eqref{eqn: NS: v} depends on $\rho$ via $p$. However, the evolution equation for $\rho$ in Eq.~\eqref{eqn: NS: rho} has now been exchanged for the constraint in Eq.~\eqref{eqn:incompressible NS constraint}.

\section{Numerical background}
\label{sec: numerical background}
The mathematical connections between the two systems suggest that numerical methods for the incompressible Navier--Stokes equations can be converted to solve QS elastoplasticity.

\subsection{First-order projection method for incompressible fluid dynamics}
\label{sec: chorin's projection method}
\begin{figure}[h!]
  \centering
  \includegraphics[width=1\textwidth]{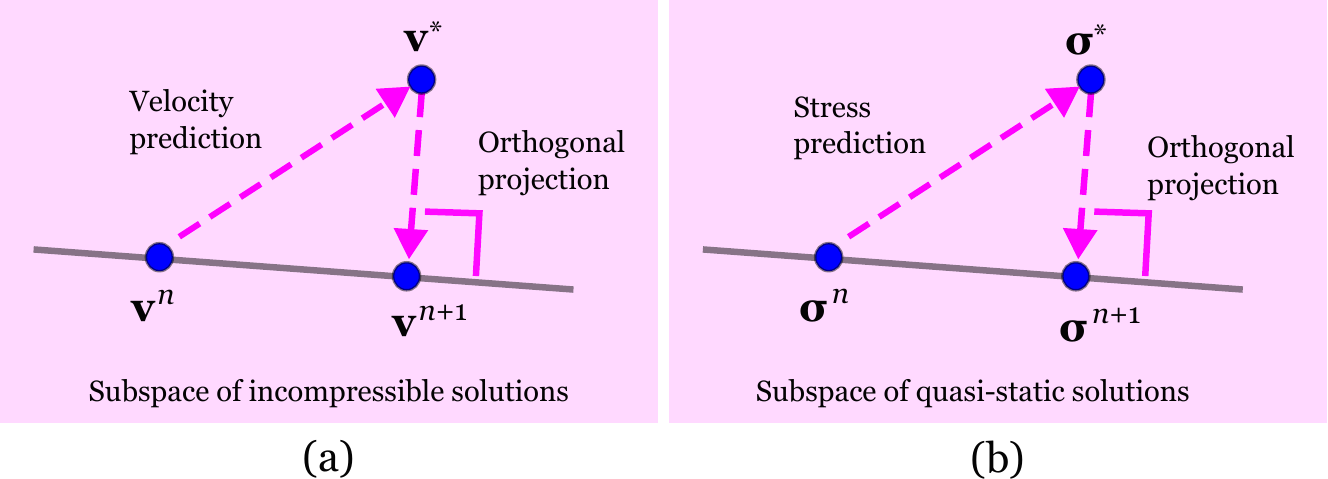}
  \caption{Schematic illustrations of (a) the projection method for incompressible Navier--Stokes equations; (b) the projection method for QS elastoplasticity. ~\label{fig:projection illustration}}
\end{figure}

In 1968, Chorin~\cite{Chorin1967,Chorin1968} devised a first-order in time projection method to solve the incompressible Navier--Stokes equations. Consider a vector space $V_{\vv}$ of all velocity fields, the divergence-free velocity field is a subspace of $V_{\vv}$. To evolve velocity $\vv_n$ at the $n^{\text{th}}$ timestep forward for $\Delta t$, to $\vv_{n+1}$ at the $(n+1)^{\text{th}}$ timestep, we can first calculate an intermediate velocity, $\vv_*$, that is not divergence-free. We can then project $\vv_*$ onto the divergence-free subspace, to solve for $\vv_{n+1}$. The idea is illustrated in the schematic plot in Fig.~\ref{fig:projection illustration}(a).

In the incompressible limit, $(\nabla \cdot \boldsymbol\tau)/\rho$ in Eq.~\eqref{eqn: NS: v} simplifies to $\nu \nabla^2 \vv$ where $\nu$ is the dynamic viscosity. To calculate $\vv_*$, we can neglect the pressure term in Eq.~\eqref{eqn: NS: v} and obtain
\begin{equation}
\label{eqn:projection:NS:intermediate velocity}
\frac{\vv_*-\vv_n}{\Delta t}=-(\vv_n\cdot\nabla)\vv_n+\nu\nabla^2 \vv_n.
\end{equation}
Comparing Eq.~\eqref{eqn:projection:NS:intermediate velocity} with Eq.~\eqref{eqn: NS: v}, we have
\begin{equation}
\label{eqn:projection:NS:intermediate velocity and velocity n+1}
\frac{\vv_{n+1}-\vv_*}{\Delta t}=-\nabla p_{n+1}.
\end{equation}
Taking the divergence of Eq.~\eqref{eqn:projection:NS:intermediate velocity and velocity n+1} and enforcing the constraint in Eq.~\eqref{eqn:incompressible NS constraint}, $\nabla\cdot \vv_{n+1}=0$, we have
\begin{equation}
\label{eqn:projection:NS:intermediate velocity and pressure n+1}
\nabla\cdot\vv_*=\Delta t\nabla^2 p_{n+1}.
\end{equation}
Equation \eqref{eqn:projection:NS:intermediate velocity and pressure n+1} is an elliptic PDE with source term given by $\nabla\cdot\vv_*$, and can be solved numerically. Substituting $p_{n+1}$ into Eq.~\eqref{eqn:projection:NS:intermediate velocity and velocity n+1} yields the solution $\vv_{n+1}$.

An inner product is also defined to establish the orthogonality of the fluid projection method. For any two velocities, $\vec{a},\vec{b}\in V_{\vv}$, their inner product is defined as
\begin{equation}
\label{eqn:projection:NS:inner product}
\langle\vec{a},\vec{b} \rangle=\int \vec{a}\cdot \vec{b} \,d^3\vx.
\end{equation}
With this definition, the projection $\vv_P=\vv_{n+1}-\vv_*$ satisfies $\langle\vv_{n+1}-\vv_n,\vv_P \rangle=0$, and therefore it is orthogonal to the divergence-free solution subspace. The notion of orthogonality ensures that the projection step projects $\vv_*$ to $\vv_{n+1}$ without introducing any additional contribution to the solution from the space orthogonal to the projection~\cite{chorin1979fluid}, which can cause an artificial drift in the solution over time.

The projection method has a splitting error due to the procedure of decomposing Eq.~\eqref{eqn: NS: v} into Eq.~\eqref{eqn:projection:NS:intermediate velocity} for the intermediate velocity $\vv_{*}$ that is not divergence-free, and into Eq.~\eqref{eqn:projection:NS:intermediate velocity and velocity n+1} for the divergence-free velocity $\vv_{n+1}$. The splitting error is $O(\Delta t)$ due to neglecting the $-\nabla p$ term in Eq.~\eqref{eqn:projection:NS:intermediate velocity}. Furthermore, the projection method uses the forward Euler scheme in Eq.~\eqref{eqn:projection:NS:intermediate velocity} to explicitly step forward in time. The forward Euler method is first-order accurate, and therefore, it has $O(\Delta t)$ timestepping error.

\subsection{First-order projection method for QS elastoplasticity}
\label{sec: first-order projection method for quasi-static elastoplasticity}
Rycroft et al.~\cite{Rycroft2015BMG2D} proposed an analogous first-order projection method to solve the QS elastoplasticity equations, drawing inspiration from the mathematical connections with the incompressible Navier--Stokes equations. Consider a vector space $V_{\bsig}$ of all stress fields, the divergence-free stress field is a subspace of $V_{\bsig}$. To evolve stress $\bsig_n$ for a timestep $\Delta t$ to $\bsig_{n+1}$, the main idea is to first solve for an intermediate stress $\bsig_*$ that is not divergence-free, then project $\bsig_*$ onto the divergence-free subspace to find the solution $\bsig_{n+1}$. The idea is shown in the schematic plot in Fig.~\ref{fig:projection illustration}(b).

First, neglecting $\bC:\bD$ in Eq.~\eqref{eqn:linear constitutive eqn}, we can solve an intermediate stress $\bsig_*$,
\begin{equation}
\label{eqn:projection:intermediate stress}
\frac{\bsig_*-\bsig_n}{\Delta t}=-\bsig_n \cdot \bome_n+\bome_n \cdot \bsig_n - (\vv_n\cdot\nabla)\bsig_n-\bC:\bD_n^\text{pl}.
\end{equation}
Comparing Eqs.~\eqref{eqn:projection:intermediate stress} \& \eqref{eqn:linear constitutive eqn}, we obtain
\begin{equation}
\label{eqn:projection:stress and CD}
\frac{\bsig_{n+1}-\bsig_*}{\Delta t}=\bC:\bD_{n+1}.
\end{equation}
Taking the divergence of Eq.~\eqref{eqn:projection:stress and CD} and enforcing the constraint in Eq.~\eqref{eqn:divergence free constraint}, we have
\begin{equation}
\label{eqn:projection:solving v at time n+1}
\nabla \cdot \bsig_*=-\Delta t \nabla \cdot (\bC:\bD_{n+1}).
\end{equation}
Since
$\mathbf{D}=(\nabla \mathbf{v}+(\nabla \mathbf{v})^{\mathsf{T}})/2$, discretizing Eq.~\eqref{eqn:projection:solving v at time n+1} using finite differences thus forms a linear system for the velocity $\vv_{n+1}$. The system can be solved via numerical linear algebra techniques such as the multigrid method~\cite{Briggs2000multigrid,Demmel1997LA}. Once $\vv_{n+1}$ is solved, we can then calculate $\bD_{n+1}$ and plug $\bD_{n+1}$ back into Eq.~\eqref{eqn:projection:stress and CD} to solve for $\bsig_{n+1}$.

Rycroft et al.~\cite{Rycroft2015BMG2D} also established an inner product to describe the orthogonality of the projection method. For two stresses $\mathbf{a},\mathbf{b}\in V_{\bsig}$, the inner product is defined as
\begin{equation}
\label{eqn:projection:inner product}
\langle\mathbf{a},\mathbf{b}\rangle=\int \mathbf{a}:\bS:\mathbf{b} \,d^3\vx,
\end{equation}
where $\bS=\bC^{-1}$, the compliance tensor that gives the infinitesimal strain $\beps$ in terms of stress, $\beps=\bS:\bsig$. The projection step is given by $\bsig_P=\bsig_{n+1}-\bsig_*$. Rycroft et al.~\cite{Rycroft2015BMG2D} showed that $\langle\bsig_{n+1}-\bsig_n, \bsig_P\rangle=0$. Therefore, the projection is orthogonal to the subspace of QS solutions, and does not cause any artificial drift in the solution over time.

The original projection method has a splitting error, which comes from decomposing Eq.~\eqref{eqn:linear constitutive eqn} into Eq.~\eqref{eqn:projection:intermediate stress} for the intermediate stress $\bsig_*$ that is not divergence-free, and into Eq.~\eqref{eqn:projection:stress and CD} for the divergence-free stress $\bsig_{n+1}$. Similar to Chorin's fluid projection method, the QS elastoplasticity projection method here has $O(\Delta t)$ splitting error due to the neglection of the $\bC:\bD$ term in Eq.~\eqref{eqn:projection:intermediate stress}. Furthermore, the method uses the forward Euler scheme in Eq.~\eqref{eqn:projection:intermediate stress} to explicitly step forward in time. The forward Euler method is first-order accurate in time, and therefore, it has $O(\Delta t)$ timestepping error.

\subsection{Plasticity model and its numerical challenges}
\label{sec: plasticity model and its numerical challenges}
To describe plasticity of the material, that is, $\bD^{\text{pl}}$ in Eq.~\eqref{eqn:linear constitutive eqn}, we use a plasticity model based on the athermal form of the shear transformation zone (STZ) theory of amorphous plasticity~\cite{Bouchbinder2007STZ,Langer2008STZ}. STZs are localized regions in the material susceptible to configuration changes. The assumption of the theory is that a population of STZs exists in an otherwise elastic material, and they are the weak spots prone to plastic deformation under externally applied mechanical work. A plastic event at an STZ site annihilates it, but may create new STZs in the process. In the plasticity model, an effective disorder temperature~\cite{Bouchbinder2011STZ,Bouchbinder2008Chi,Loi2008Chi,Cugliandolo2011Chi}, $\chi$, is used to measure the density of STZs in the material, and it describes the disorderliness of the system at the atomic scale. The plasticity model is described in detail in \ref{appendix: plasticity model}.

The plastic deformation tensor, $\bD^\text{pl}$, is proportional to the deviatoric stress tensor, and is written as
\begin{equation}
\label{eqn:Dpl tensor equation}
\bD^\text{pl}=\frac{\bsig_0}{\Bar{s}}D^\text{pl},
\end{equation}
where $D^\text{pl}$ is a scalar function of $\chi$ and $\Bar{s}$. When $\Bar{s}<s_Y$, $D^\text{pl}=0$. When $\Bar{s}\geq s_Y$, $D^{\text{pl}}$ is given by the plasticity model, and increases as $\chi$ increases. Equation \eqref{eqn:Dpl tensor equation} is an example of \textit{rate-dependent} plasticity model, where the stress may exceed the yield stress, and the rate of plastic deformation is determined by the specific amount of stress. This is in contrast to \textit{rate-independent} plasticity models (such as that employed by Wilkins~\cite{Wilkins1963}), where the material has a yield surface in stress space. In this case, plastic deformation ensures that the stress can never exceed yield and go beyond this surface.

The effective temperature $\chi$ responds to the plastic deformation and satisfies 
\begin{equation}
\label{eqn:Chi evolution}
c_0\frac{d\chi}{dt}=\frac{(\bD^{\text{pl}}:\bsig_0)(\chi_{\infty}-\chi)}{s_Y},
\end{equation}
where $\chi_{\infty}$ is the steady state effective temperature, and $c_0$ is the effective heat capacity; their values are given in \ref{appendix: plasticity model}. Under externally applied mechanical work, STZs are created and annihilated proportionally, and $\bD^{\text{pl}}:\bsig_0$ describes the energy dissipation rate of this process. Furthermore, as seen from Eq.~\eqref{eqn:Chi evolution}, an increase in plastic deformation $\bD^\text{pl}$ increases $\chi$, until it saturates at $\chi_{\infty}$. On the other hand, an increase in $\chi$ increases $D^\text{pl}$, and therefore increases $\bD^\text{pl}$ in Eq.~\eqref{eqn:Dpl tensor equation}. This mutual feedback of the plasticity model typically leads to shear banding~\cite{Manning2007STZ,Manning2009STZ}.

The original projection method uses the forward Euler scheme in Eq.~\eqref{eqn:Chi evolution} to explicitly step forward in time, which has $O(\Delta t)$ timestepping error. The calculation of the plastic deformation, $D^{\text{pl}}$, poses additional numerical challenges. $D^{\text{pl}}$ grows rapidly when $\Bar{s}>s_Y$. In a timestep $\Delta t$, it can induce very large deviatoric stress change $\Delta \Bar{s}$, significantly overshooting the yield surface. This can cause the forward Euler scheme of the stress update in Eq.~\eqref{eqn:projection:intermediate stress} to lose accuracy quickly. Therefore, instead of advancing for a full timestep $\Delta t$ directly, the original projection method uses an adaptive subtepping procedure, dividing $\Delta t$ into many smaller substeps. It uses a fixed threshold $\eta$ to bound the changes in the deviatoric stress, ensuring $\Delta \Bar{s}<\eta$ in each substep. The adaptive substepping scheme works well in practice but it complicates the error analysis and it is unclear how to choose $\eta$ apart from empirical testing. In this work, we therefore replace this approach with an adaptive Runge--Kutta scheme (\ref{sec: RK21FSAL}) that is more accurate, and allows for systematic error control.

\section{Second-order formulation of the projection method}
\label{sec: second-order formulation}

\subsection{Second-order projection method for fluid dynamics}
\label{sec: Second-order projection method for fluid dynamics}
Over the past fifty years many improvements have been made to Chorin's projection method~\cite{BROWN2001projection}. One general version of the second-order fluid projection method uses a fractional step procedure with an incremental pressure term $q$~\cite{bell1989projection2nd,bell1991projection2nd,goda1979projection2nd,Kim1985projection2nd,vanKan1986projection2nd}. The procedure is as follows~\cite{BROWN2001projection}:

\begin{steps}
\item Solve the intermediate velocity $\vv_*$. Compared to Eq.~\eqref{eqn:projection:NS:intermediate velocity} that calculates $\vv_*$ by ignoring the pressure term $p$ completely, here, an incremental pressure term $q$ is used, where $q$ approximates pressure at half timestep, $p_{n+1/2}$. The update equation is
    \begin{equation}
    \label{eqn:NS,projection,2nd order, intermediate velocity}
    \frac{\vv_*-\vv_n}{\Delta t}=-\nabla q -[(\vv\cdot\nabla)\vv]_{n+1/2} + \frac{\nu}{2}\nabla^2 (\vv_*+\vv_n),
    \end{equation}
    where $[(\vv\cdot\nabla)\vv]_{n+1/2}$ is a second-order approximation of the advective term evaluated at the half time point $t^{n+1/2} = t_n + \tfrac12 \Delta t$. Boundary conditions
    \begin{equation}
    \label{eqn:NS,projection,2nd order, intermediate velocity, BC}
    B(\vv_*)=0
    \end{equation}
    are applied on $\vv_*$.
  \item Perform the projection step. Comparing Eq.~\eqref{eqn:NS,projection,2nd order, intermediate velocity} with Eq.~\eqref{eqn: NS: v}, we have
    \begin{equation}
    \label{eqn:NS,projection,2nd order, v n+1 and *}
    \frac{\vv_{n+1}-\vv_*}{\Delta t}=-\nabla\Phi_{n+1}
    \end{equation}
    where $\Phi_{n+1}$ will be used to correct $q$ for approximating $p_{n+1/2}$. Taking the divergence of Eq.~\eqref{eqn:NS,projection,2nd order, v n+1 and *} and imposing the incompressibility constraint in Eq.~\eqref{eqn:incompressible NS constraint}, $\nabla \cdot \vv_{n+1}=0$, we have,
    \begin{equation}
    \label{eqn:NS,projection,2nd order, v n+1 and Phi}
    \nabla\cdot\vv_*=\Delta t \nabla^2\Phi_{n+1}.
    \end{equation}
    We can solve the elliptic PDE in Eq.~\eqref{eqn:NS,projection,2nd order, v n+1 and Phi} for $\Phi_{n+1}$.
  \item Update the pressure according to
    \begin{equation}
    \label{eqn:NS,projection,2nd order, pressure update}
    p_{n+1/2}=q+L(\Phi_{n+1}),
    \end{equation}
    where $L$ is a function representing the dependence of $p_{n+1/2}$ on $\Phi_{n+1}$. The updated pressure $p_{n+1/2}$ is then used in the next timestep as $q$.
\end{steps}
Here, the use of incremental pressure $q$ reduces the splitting error to be $O(\Delta t ^2)$. The time updating scheme in Eq.~\eqref{eqn:NS,projection,2nd order, intermediate velocity} is a semi-implicit Crank--Nicolson-type scheme~\cite{crank1947CNmethod}, which is second-order accurate in time. Therefore, the procedure gives second-order temporal accurate solution for $\vv$. Depending on the choice of the pressure correction function $L$ in Eq.~\eqref{eqn:NS,projection,2nd order, pressure update}, $p$ can achieve second-order temporal accuracy as well~\cite{BROWN2001projection}.

\subsection{Second-order projection method for QS elastoplasticity}
\label{sec:Second-order projection method for quasi-static elastoplasticity}
Inspired by the second-order fluid projection method above, we formulate an analogous second order projection method for QS elastoplasticity, using an incremental velocity term $\vq$ that approximates velocity at half timestep, $\vv_{n+1/2}$. Here, our variables of interest are $\bsig$, $\chi$, and $\vv$. To advance the variables from $(n)^{\text{th}}$ to $(n+1)^{\text{th}}$ timestep over a fixed time interval $\Delta t$, the second-order projection is as follows:

    \begin{steps}
    \item\label{stage1step1.timestepping} Solve for the intermediate stress $\bsig_*$. An incremental velocity term, $\vq$, is used to approximate $\vv_{n+1/2}$. Rather than omitting the term $\bC:\bD$ as in the first-order projection method in Eq.~\eqref{eqn:projection:intermediate stress}, $\bC:\bD_{\vq}$ is used to approximate it, where $\bD_{\vq}=\tfrac12 (\nabla \vq+(\nabla\vq)^\trans)$. We can calculate the intermediate stress $\bsig_*$ by updating $\bsig_n$ for a fixed time interval via the equation
        \begin{equation}
        \label{eqn:2nd order, intermediate stress, stage 1}
        (\bsig)_t=-\bsig_n\cdot \bome_n+\bome_n\cdot\bsig_n-(\vv_n\cdot\nabla)\bsig_n+\bC:\bD_{\vq}-\bC:\bD^{\text{pl}}(\bsig, \chi),
        \end{equation}
        where $\bD^{\text{pl}}(\bsig, \chi)$ is a function dependent on $\bsig$ and $\chi$, and it represents the plastic deformation given by the STZ plasticity model. The advection and spin terms at time $n$, along with $\bC:\bD_{\vq}$, are used in the equation, and the advection term is calculated using the second-order ENO numerical scheme as given in Eq.~\eqref{eqn:corner advective derivative}. These terms are therefore known at the beginning of the timestep. They can be viewed as constants, serving as source terms to Eq.~\eqref{eqn:2nd order, intermediate stress, stage 1}. Therefore, Eq.~\eqref{eqn:2nd order, intermediate stress, stage 1} is effectively an ODE of stress evolution for the new timestep.
        
        In addition, Eq.~\eqref{eqn:2nd order, intermediate stress, stage 1} is coupled with the evolution of $\chi$, and we solve for the effective temperature at the $(n+1)^{\text{th}}$ timestep, $\chi_{n+1}$, evolving from the previous timestep $\chi_{n}$ via
        \begin{equation}
          \label{eqn:2nd order, chi, stage 1}
          (\chi)_t=-(\vv_n\cdot\nabla)\chi_n+F(\bsig, \chi)
        \end{equation}
        where $F(\bsig, \chi)$ is a function dependent on $\bsig$ and $\chi$. It is given by the STZ model and represents the effect of the plastic deformation on the $\chi$ field. The advection term at time $n$, discretized using the second-order ENO scheme, is used as a source term for the PDE. Therefore, Eq.~\eqref{eqn:2nd order, chi, stage 1} is effectively an ODE of $\chi$ evolution for the new timestep.

        To solve $\bsig_*$ and $\chi_{n+1}$ from the coupled ODE system of Eqs.~\eqref{eqn:2nd order, intermediate stress, stage 1} \& \eqref{eqn:2nd order, chi, stage 1}, and achieve second order temporal accuracy, we use an efficient explicit Runge--Kutta (RK) 2(1) first-same-as-last (FSAL) adaptive timestepping scheme~\cite{ralston1965,runge1895RKmethod,kutta1901RKmethod}, and control the local timestepping error to be of order $O(\Delta t^2)$. The implementation of RK2(1) and its error control are detailed in \ref{sec: RK21FSAL}.

      \item\label{stage1step2.projection} Perform the projection. By comparing Eq.~\eqref{eqn:2nd order, intermediate stress, stage 1} and Eq.~\eqref{eqn:linear constitutive eqn}, we obtain
        \begin{equation}
        \label{eqn:2nd order, projection, stress n+1, *, stage 1}
        \frac{\bsig_{n+1}-\bsig_*}{\Delta t}=\bC:\bD(\bPhi_*),
        \end{equation}
        where $\bD(\bPhi_*)=\tfrac12(\nabla\bPhi_* +(\nabla\bPhi_*)^\trans)$. Taking the divergence of Eq.~\eqref{eqn:2nd order, projection, stress n+1, *, stage 1} and imposing the QS constraint given by Eq.~\eqref{eqn:divergence free constraint}, $\nabla\cdot\bsig_{n+1}=0$, we have
        \begin{equation}
        \label{eqn:2nd order, projection, div stress *, stage 1}
        \nabla\cdot\bsig_*=-\Delta t \nabla\cdot(\bC:\bD(\bPhi_*)).
        \end{equation}
        Section~\ref{sec: FEM projection} describes an FEM projection method to solve Eq.~\eqref{eqn:2nd order, projection, div stress *, stage 1} for the velocity correction term $\bPhi_*$. Using $\bPhi_*$, we then correct $\vq$ to better approximate $\vv_{n+1/2}$.
      \item Correct $\vq$ to calculate a more accurate approximation of $\vv_{n+1/2}$, $\hat{\vv}_{n+1/2}$. Then update $\vq$ to $\hat{\vv}_{n+1/2}$, 
        \begin{equation}
        \label{eqn:2nd order, projection, q correction, stage 1}
        \begin{split}
            \hat{\vv}_{n+1/2}&=\vq+\bPhi_*, \\
            \vq&=\hat{\vv}_{n+1/2}.
        \end{split}
        \end{equation}
        $\vq$ is then used in the next timestep as the incremental velocity.
        Furthermore, we update $\vv_{n+1}$ as
        \begin{equation}
        \label{eqn:2nd order, projection, velocity n+1 update, stage 2}
        \vv_{n+1}=\vq.
        \end{equation}
        Using the relationship given by Eq.~\eqref{eqn:2nd order, projection, stress n+1, *, stage 1}, compute $\bsig_{n+1}$
        \begin{equation}
        \label{eqn:2nd order, projection, stress n+1 update, stage 2}
        \bsig_{n+1}=\bsig_*+\Delta t \bC:\bD(\bPhi_*).
        \end{equation}
\end{steps}
  
As mentioned in Sec.~\ref{sec: numerical background}, there are several error sources in the original projection method that contributed to its first-order temporal accuracy. Here, we discuss how these errors have been reduced to second-order temporal accuracy using the projection procedure with incremental velocity $\vq$. By using an adaptive RK2(1) timestepping scheme in \ref{stage1step1.timestepping}, we control the timestepping error to be of order $O(\Delta t^2)$, improving upon the original first-order forward Euler method as described in Sec.~\ref{sec: first-order projection method for quasi-static elastoplasticity} and Sec.~\ref{sec: plasticity model and its numerical challenges}.

Furthermore, the adaptive RK2(1) scheme resolves the numerical difficulty of the plasticity related terms, $D^{\text{pl}}$ and $F$. As mentioned in Sec.~\ref{sec: plasticity model and its numerical challenges}, when $\Bar{s}>s_Y$, $D^{\text{pl}}$ grows rapidly and can cause inaccuracy of the forward Euler scheme. Here, using RK2(1), we can approximate the order of errors in a local timestep, and we can relate the adaptive timestepping rules to criteria on the solution errors. Through this adaptive timestepping process, the calculation of the plasticity terms are resolved automatically, maintaining second order temporal accuracy.

Lastly, the use of the incremental velocity $\vq$ reduces the splitting error of the projection scheme to be $O(\Delta t^2)$. The incremental velocity $\vq$ is an improvement from the original formulation, since now we have a term $\bC:\bD_{\vq}$ to approximate $\bC:\bD$ in the timestepping, rather than omitting the term altogether as in Sec.~\ref{sec: first-order projection method for quasi-static elastoplasticity}.  Combining all of the above, the numerical scheme has second order temporal accuracy for all fields of interest, $\bsig$, $\chi$, and $\vv$.

\section{Finite element formulation of the projection step}
\label{sec: FEM projection}
We derived an FEM~\cite{donea_fluid_fem,claes_fem} formulation to solve the projection step for both the original first-order projection method and the improved second-order projection method.  For the original projection method, we aim to solve for $\vv_{n+1}$ using Eq.~\eqref{eqn:projection:solving v at time n+1}. The FEM velocity boundary condition is $\vv^B=(u^B,v^B)$, given by the Dirichlet boundary condition on the upper and lower walls of the 2D simple shear simulation, as described in Sec.~\ref{sec: set up of numerical tests}. The weak form formulation of the FEM procedure for the original projection method is provided in Sec.~\ref{sec:Derivation of the FEM projection step}. 
The FEM projection gives rise to a linear system to solve, $\bA \vw = \vb$; The choice of element functions and the derivation of the component forms of the linear system are discussed in \ref{appendix:FEM linear system details} in the supplementary material.

For the second-order projection method, where we aim to solve for the correction velocity $\bPhi_*$ via Eq.~\eqref{eqn:2nd order, projection, div stress *, stage 1}, the FEM derivation procedure is exactly the same, except that we replace the FEM velocity boundary condition to be $\bPhi_B=\mathbf{0}$. This is because in Eq.~\eqref{eqn:2nd order, projection, q correction, stage 1}, $\vq$ is corrected by $\vq=\vq+\bPhi_*$. Since $\vq$ satisfies the velocity Dirichlet boundary conditions $\vq^B=(u^B,v^B)$ at all time, the correction velocities on the boundaries are $\bPhi_B=\mathbf{0}$. The FEM linear system associated with the second-order projection method in component form is briefly discussed in \ref{appendix: FEM second-order projection}.

\subsection{Derivation of the FEM projection step}
\label{sec:Derivation of the FEM projection step}
To derive the finite-element method, we first convert the original PDE system into the equivalent weak formulation. We denote $\mathcal{L}_2(\Omega)$ as the space of square integrable functions over the 2D domain $\Omega$. The space is equipped with the standard inner product and norm, respectively,
\begin{equation}
\label{eqn:fem projection, L2, inner product}
(v,w)=\int_{\Omega}vw \,d\vx, \quad \lVert v \rVert=(v,v)^{\frac{1}{2}}.
\end{equation}
The scalar Hilbert space $\mathcal{H}^1(\Omega)$ is defined as
\begin{equation}
\label{eqn:fem projection, hilbert space H10}
\mathcal{H}^1(\Omega)=\{v: v\in \mathcal{L}_2(\Omega), \ \frac{\p v}{\p x}\in L_2(\Omega),\  \frac{\p v}{\p y}\in L_2(\Omega)\}.
\end{equation}
The space has inner product and norm,
\begin{equation}
\label{eqn:fem projection, hilbert space H10, inner product}
(v,w)_{\mathcal{H}^1(\Omega)}=\int_{\Omega}(vw+\nabla v \cdot \nabla w)\,d\vx, \quad \lVert v \rVert_{\mathcal{H}^1(\Omega)} =\sqrt{(v,v)_{\mathcal{H}^1(\Omega)}}.
\end{equation}
Since we are solving for velocity $\vv$, we extend to the space of vector functions with two components, $\vv=(v_1,v_2)$. We define the space $\bmcL_2(\Omega)$ as a space where each component of the vector functions $v_1, v_2\in \mathcal{L}_2(\Omega)$. It has inner product and norm defined by
\begin{equation}
\label{eqn:fem projection, vector space L2, inner product}
(\vv,\vw)_{\bmcL_2(\Omega)}=\int_{\Omega} \vv \cdot \vw \,d\vx, \quad
\lVert \vv \rVert_{\bmcL_2(\Omega)} =\sqrt{\lVert v_1 \rVert^2+\lVert v_2 \rVert^2}.
\end{equation}
We define the space $\bmcH^1(\Omega)$, the space of vector functions for which each component $v_1, v_2\in \mathcal{H}^1(\Omega)$. It has inner product and norm defined by
\begin{equation}
\label{eqn:fem projection, vector hilbert space H10, inner product}
(\vv,\vw)_{\bmcH^1(\Omega)}=\int_{\Omega}(\vv \cdot \vw+\nabla \vv : \nabla \vw)\,d\vx, \quad
\lVert \vv \rVert_{\bmcH^1(\Omega)} =\sqrt{\lVert v_1 \rVert_{H^1(\Omega)}^2+\lVert v_2 \rVert_{H^1(\Omega)}^2}.
\end{equation}
We have Dirichlet boundary conditions on the upper and lower walls for this problem, and we are solving for the unknown velocities inside the domain. To impose the Dirichlet boundary conditions in the formulation, we define our test function space, $\bmcH_0^1(\Omega)$, by restricting $\bmcH^1(\Omega)$ to functions that vanish on the domain boundary $\Gamma$,
\begin{equation}
\label{eqn:fem projection, hilbert space H10}
\bmcH_0^1(\Omega)=\{\vv\in \bmcH^1(\Omega) | \vv=\mathbf{0} \text{ on } \Gamma \}.
\end{equation}
We also define a space for trial solution functions. This collection of functions is similar to the test functions, except that they are required to satisfy the Dirichlet conditions on $\Gamma$. We denote the space by $\bmcS_B$, and it is defined by
\begin{equation}
\label{eqn:fem projection, trial solution space S}
\bmcS_B(\Omega)=\{\vv\in \bmcH^1(\Omega) | \vv=\vv^B \text{ on } \Gamma \}.
\end{equation}
Let $\vLambda(\vx)\in \bmcH_0^1(\Omega)$ be a test solution we choose. We can multiply $\vLambda(\vx)$ on both sides of Eq.~\eqref{eqn:projection:solving v at time n+1}, and take the integral in the simulation domain $\Omega$:
\begin{equation}
\label{eqn:fem projection, main}
\int_{\Omega}\vLambda(\vx)\nabla\cdot\bsig_{*}\,d\vx=-\Delta t \int_{\Omega}\vLambda(\vx)\nabla\cdot(\bC:\bD_{n+1})\,d\vx.
\end{equation}
Note that $\bsig_*$ and $\bC:\bD_{n+1}$ are both symmetric $2$-tensors. The components of $\bsig_*$ are given by Eq.~\eqref{eqn:stress tensor component}. In component form $\bC:\bD$ is
\begin{equation}
\label{eqn:CD tensor component}
\bC:\bD=
\begin{pmatrix}
(K+\frac{4}{3}\mu)\frac{\p u}{\p x} +(K-\frac{2}{3}\mu)\frac{\p v}{\p y} & \mu (\frac{\p v}{\p x}+\frac{\p u}{\p y}) \\
\mu (\frac{\p v}{\p x}+\frac{\p u}{\p y}) & (K-\frac{2}{3}\mu)\frac{\p \mu}{\p x}+(K+\frac{4}{3}\mu)\frac{\p v}{\p y} \\
\end{pmatrix},
\end{equation}
Using Green's first identity on vector field, let $\vn$ be the outward pointing unit normal to the boundary $\Gamma$. The left hand side of Eq.~\eqref{eqn:fem projection, main} becomes
\begin{equation}
\label{eqn:fem projection, main, LHS}
\int_{\Omega}\vLambda(\vx)\nabla\cdot\bsig_{*}\,d\vx
=\int_{\Gamma}\vLambda(\vx)(\bsig_*\cdot \vn)\,d\vx-\int_{\Omega}\bsig_*: \nabla\vLambda \,d\vx =-\int_{\Omega}\bsig_*: \nabla\vLambda \,d\vx,
\end{equation}
where the boundary integral disappears because $\vLambda(\vx)$ comes from a function space $\bmcH_0^1(\Omega)$ that vanishes on $\Gamma$.
Using Green's first identity on vector field for the integral on the right hand side of Eq.~\eqref{eqn:fem projection, main}, we obtain
\begin{equation}
\label{eqn:fem projection, main, RHS}
\begin{aligned}
\int_{\Omega}\vLambda(\vx)\nabla\cdot(\bC:\bD_{n+1})d\vx
&=\int_{\Gamma}\vLambda(\vx)((\bC:\bD_{n+1})\cdot \vn)d\vx-\int_{\Omega}(\bC:\bD_{n+1}):\nabla\vLambda(\vx)\,d\vx \\
&=-\int_{\Omega}(\bC:\bD_{n+1}):\nabla\vLambda(\vx)d\vx,
\end{aligned}
\end{equation}
where the boundary integral disappears because $\vLambda(\vx)$ vanishes on $\Gamma$. Therefore, substituting Eqs.~\eqref{eqn:fem projection, main, LHS} \& \eqref{eqn:fem projection, main, RHS} into Eq.~\eqref{eqn:fem projection, main}, we have
\begin{equation}
\label{eqn:fem projection, main, rearrange keeping CD whole}
\begin{aligned}
\int_{\Omega}(\bC:\bD_{n+1}):\nabla\vLambda(\vx)\,d\vx
= -\frac{1}{\Delta t}  \int_{\Omega}\bsig_* : \nabla\vLambda \,d\vx.
\end{aligned}
\end{equation}

We aim to solve for the velocities $\vv_{n+1}$ by discretizing Eq.~\eqref{eqn:fem projection, main, rearrange keeping CD whole}. To do so, we first choose a finite-dimensional subspace $\bmcV_h(\Omega)\in \bmcH_0^1(\Omega)$, according to the simulation grid spacing $h$. We take $\bmcV_h$ as a space of functions that, in each of the $x$ and $y$ dimensions, are piecewise linear on the grid cells of our simulation domain and vanish on the boundary. In each dimension, we can choose the same nodal basis functions $\psi$, defined on each inner node in the domain. Then we can choose the nodal basis functions $\vpsi$ for $\bmcV_h$ to have the forms $(\psi,0)$ for the $x$ dimension, and $(0,\psi)$ for the $y$ dimension, defined on each inner node $I$. Similarly, we can choose a finite-dimensional subspace $\bmcS_h(\Omega)\in\bmcS_B(\Omega)$ in the same way, using the same type of piecewise linear nodal basis functions in each dimension, except that the functions satisfy the Dirichlet boundary conditions on $\Gamma$. Let $\mathcal{V}_h(\Omega)$ denote the scalar finite-dimensional subspace in each dimension of $\bmcV_h$, and let $\mathcal{S}_h(\Omega)$ denote the ones for $\bmcS_h$.

It is convenient to separate the $x$ and $y$ dimensions at this point for further derivation of the FEM scheme. Let $k=1,2$ represent the dimensions. For our simple shear simulation, the top and bottom wall velocities are known and imposed as Dirichlet boundary conditions. Suppose the total number of grid cells in our simulation grid is $\mathcal{M}=MN$, and there are $\mathcal{B}$ Dirichlet boundary nodes and $\mathcal{I}$ inner nodes. Denote the known boundary velocities as $\vv^B=(u^B, v^B)=(v_1^B,v_2^B)$, and the unknown velocities at the inner nodes as $\vv^I_{n+1}=(u_{n+1}^I, v_{n+1}^I)=(v_{n+1,1}^I, v_{n+1,2}^I)$. The unknown $v_{n+1,k}^I$ can be expressed as sum of nodal basis functions of $\mathcal{V}_h$ with unknown weights $w_{i,k}, i=1,\ldots,\mathcal{I}$, and the known $v_k^B$ can be expressed as sum of nodal basis functions on $\Gamma$ of $\mathcal{S}_h$,
\begin{equation}
\label{eqn:fem projection, u and v n+1 expressed as basis functions}
\begin{aligned}
    v_{n+1,k}^I
    &=\sum_{i=1}^{\mathcal{I}} w_{i,k}\psi_{i}(\vx), \quad \psi_{i}(\vx)\in \mathcal{V}_h, \\
    v_{k}^B
    &=\sum_{j=1}^{\mathcal{B}} v_k^B\psi_{j}(\vx), \quad v_k^B\psi_{j}(\vx)\in \mathcal{S}_h,
    \quad k=1,2.
\end{aligned}
\end{equation}
For the integral on the left hand side of Eq.~\eqref{eqn:fem projection, main, rearrange keeping CD whole}, $\bD_{n+1}$ can be split into two terms, $\bD_{n+1}=\bD^B+\bD^I_{n+1}$, where
\begin{equation}
\label{eqn:fem projection, CD_B and CD_I expressed in velocities}
\mathbf{D}^{B}=
\frac{\nabla \mathbf{v}^{B}+(\nabla \mathbf{v}^{B})^{\mathsf{T}}}{2}, \quad
\mathbf{D}^{I}_{n+1}=
\frac{\nabla \mathbf{v}_{n+1}^{I}+(\nabla \mathbf{v}_{n+1}^{I})^{\mathsf{T}}}{2}.
\end{equation}
Then Eq.~\eqref{eqn:fem projection, main, rearrange keeping CD whole} becomes
\begin{equation}
\label{eqn:fem projection, main, rearrange CD = CD_B+CD_I}
\begin{aligned}
\int_{\Omega}(\bC:\bD_{n+1}^I):\nabla\vLambda(\vx)\,d\vx
= -\int_{\Omega}(\bC:\bD^B):\nabla\vLambda(\vx)\,d\vx
-\frac{1}{\Delta t}  \int_{\Omega}\bsig_*: \nabla\vLambda \,d\vx.
\end{aligned}
\end{equation}
We restrict the test function $\vLambda(\vx)=(\Lambda_1(\vx),\Lambda_2(\vx))$ to be the nodal basis functions $\vpsi$ of $\bmcV_h(\Omega)$. Therefore, for $\vLambda=(\psi,0)$ and $\vLambda=(0,\psi)$, we have, respectively,
\begin{equation}
\label{eqn:fem projection, derivative of nodal basis dimension 1 and 2}
\nabla\vLambda=
  \begin{bmatrix}
    \p \psi /\p x  & \p \psi /\p y \\
    0 & 0
  \end{bmatrix};
  \quad
  \nabla\vLambda=
  \begin{bmatrix}
  0 & 0 \\
    \p \psi /\p x  & \p \psi /\p y
  \end{bmatrix},
\end{equation}
where each picks out the $k^{\text{th}}$ row of the $2$-tensors $\bC:\bD_{n+1}^I$, $\bC:\bD^B$ and $\bsig_*$ in Eq.~\eqref{eqn:fem projection, main, rearrange CD = CD_B+CD_I}. Let $(\bC:\bD)_k$ and $\bsig_k$ represent the $k^{\text{th}}$ row of the $2$-tensors; note that since they are all symmetric tensors, the entries in the $k^{\text{th}}$ row are the same as the entries in the $k^{\text{th}}$ column. Putting the unknown terms to the left-hand side and separating the dimensions, and rewriting using vector--vector inner products, Eq.~\eqref{eqn:fem projection, main, rearrange CD = CD_B+CD_I} is equivalent to
\begin{equation}
\label{eqn:fem projection, main, rearrange}
\begin{aligned}
\int_{\Omega}(\bC:\bD_{n+1}^I)_k\cdot\nabla\psi(\vx)\,d\vx
&=
-\int_{\Omega}(\bC:\bD^B)_k\cdot\nabla\psi(\vx)\,d\vx
-
\frac{1}{\Delta t}  \int_{\Omega}\bsig_{*,k}\cdot \nabla\psi \,d\vx  , \\
\nabla\psi&=
\begin{bmatrix}
    \p \psi/ \p x \\ \p \psi/ \p y
\end{bmatrix}, \quad k=1,2.
\end{aligned}
\end{equation}

Next we can discretize Eq.~\eqref{eqn:fem projection, main, rearrange} by representing $\bD_{n+1}^I$ and $\bD^B$ using the nodal basis functions of $\bmcV_h$ and $\bmcS_h$ via Eq.~\eqref{eqn:fem projection, CD_B and CD_I expressed in velocities} \& \eqref{eqn:fem projection, u and v n+1 expressed as basis functions}, and varying the test function in each dimension $k=1,2$ to be $\{\psi_{i}\}_{i = 1, \ldots,\mathcal{I}}$. Then we obtain a linear system $\bA\vw=\vb$ of $2\mathcal{I}$ equations, to solve for the $2\mathcal{I}$ unknown inner node velocity weights, $\{w_{i,k}\}_{i = 1, \ldots,\mathcal{I}; k=1,2}$. Here, $\vb$ is given by the right-hand side of Eq.~\eqref{eqn:fem projection, main, rearrange}. $\bA\vw$ is given by the left-hand side of Eq.~\eqref{eqn:fem projection, main, rearrange}, where the inner node velocity weights make up the entries of $\vw$.

Details of the derivation of the linear system in component form with regard to the nodal basis functions of $V_h$ are provided in \ref{appendix:FEM linear system details}. The derivation in \ref{appendix:FEM linear system details} is for the original first-order projection method. For the improved second-order projection method, the FEM formulation of the projection step follows similar derivation. Its linear system in component form is briefly discussed in \ref{appendix: FEM second-order projection}.

\subsection{Advantages of the FEM projection step}
\label{sec:Advantages of the FEM projection step}
The matrix $\bA$ from the FEM formulation is sparse and symmetric positive-definite (SPD), which provides many benefits in solving the linear system. For example, both the Gauss--Seidel~\cite{Seidel1874} and conjugate gradient methods~\cite{hestenes1952CG,straeter1971Minimization} are guaranteed to converge for SPD matrices. The boundary conditions can be set up easily in FEM. The Neumann boundary condition can be naturally implemented in the derivation of the FEM formulation. The Dirichlet boundary condition can be easily implemented by restricting the test function space $\bmcH^1$ to a subspace $\bmcH^1_0$ that satisfies the Dirichlet condition. In comparison, in the original scheme~\cite{Rycroft2015BMG2D}, the linear system comes directly from the finite different discretization of Eq.~\eqref{eqn:projection:solving v at time n+1}. The matrix $\bA$ in that case is not guaranteed to be symmetric for the boundary terms. In addition, the boundary conditions, especially the Neumann boundary condition, need to be exactly set up in the linear system, which may be complicated to do.

\section{Adaptive global timestepping}
\label{sec:Adaptive global timestepping}
We developed an adaptive global time-stepping procedure, by bounding the projection step size $\lVert\bsig_P\rVert=\lVert\bsig_{n+1}-\bsig_{*}\rVert$. The main idea is as follows: we can use the inner product defined in Eq.~\eqref{eqn:projection:inner product} to measure the projection step size $\lVert\bsig_P\rVert$ of our second-order projection method described in Sec.~\ref{sec:Second-order projection method for quasi-static elastoplasticity}. The projection step size represents how close the system is to being quasi-static. We can develop an adaptive global timestepping procedure for the second-order projection method, by setting a maximum tolerance of the projection step size. The procedure then allows us to achieve high accuracy of the solutions with significantly fewer timesteps.

\subsection{Measuring the size of the projection}
\label{sec:Measuring the size of the projection}
To measure the projection size of the improved projection method described in Sec.~\ref{sec:Second-order projection method for quasi-static elastoplasticity}, we can calculate $\bsig_P=\bsig_{n+1}-\bsig_*$ from \ref{stage1step2.projection}. We can then compute $\lVert\bsig_P\rVert$ for the adaptive global timestepping scheme. We aim to bound the amount of projection $\lVert\bsig_P\rVert$, so that the system remains close to quasi-staticity.

For a simulation domain $\Omega$ and a timestep of size $\Delta t$, we perform a projection $\bsig_P$ in \ref{stage1step2.projection}. Using the inner product defined in Eq.~\eqref{eqn:projection:inner product}, an appropriate measure of quasi-staticity is
\begin{equation}
\label{eqn:global adaptive timestepping, Q calculation}
Q=\frac{\sqrt{\langle \bsig_P, \bsig_P \rangle}}{\Delta t}
\end{equation}
and therefore,
\begin{equation}
\label{eqn:global adaptive timestepping, Q^2 calculation}
Q^2=\frac{1}{\Delta t^2}\int_{\Omega} ((3\lambda+2\mu)\bsig_P:\bsig_P-\lambda(\Tr \bsig_P)^2) d^2 \vx.
\end{equation}
In terms of the components of $\bsig_P$, we have
\begin{equation}
\label{eqn:global adaptive timestepping, Q^2 calculation, component of s_P}
Q^2=\frac{6}{\Delta t^2}\int_{\Omega} (K\Bar{s}_P^2+\mu p_P^2) d^2 \vx
\end{equation}
where $\Bar{s}_P$ and $p_P$ are the magnitude of deviatoric stress and the pressure component of $\bsig_P$, respectively. A quadrature rule is used to evaluate the integral on $\Omega$ for $Q^2$, and then we take the square root to obtain $Q$. We further scale $Q$ with the area of the domain $|\Omega|$ and the shear modulus $\mu$, to obtain a dimensionless measure of quasi-staticity,
\begin{equation}
\label{eqn:global adaptive timestepping, Q_dimensionaless calculation}
\hat{Q}=\frac{Q}{|\Omega|\mu}.
\end{equation}
We then further compute a measure of the amount of projection that happened in the timestep,
\begin{equation}
\label{eqn:global adaptive timestepping, Q* computation}
Q^*=\hat{Q}\Delta t.
\end{equation}
We bound $Q^*$ in our adaptive global timestepping procedure.

\subsection{Adaptive global timestepping criteria and procedure}
\label{sec:Adaptive global timestepping criteria and procedure}
We start with an initial timestep of size $\Delta t_{\text{init}}$. For a simulation with grid spacing $\Delta x = \Delta y = h$, in each simulation timestep, we use a tolerance $Q^*_{\text{tol}}$ to bound $Q^*$, and $Q^*_{\text{tol}}=F_Q h^2$, where $F_Q$ is a scalar factor. We chose $Q^*_{\text{tol}}\propto h^2$, since the simulation has $O(h^2)$ spatial discretization error, and therefore the final solution accuracy can at best be $O(h^2)$. We link the projection errors to the spatial discretization error, so that it is comparable in size to $O(h^2)$.

In a current timestep of size $\Delta t_n$, we compute the next timestep $\Delta t_{n+1}$ as
\begin{equation}
\label{eqn:global adaptive timestepping, new timestep calculation}
\Delta t_{n+1}=\min\{ \Delta t_{\text{max}},   \max\{\Delta t_{\text{min}}, \Delta t_{n} \min \{F_{\text{max}}, F_s\cdot\sqrt{Q^*_{\text{tol}}/Q^*} \} \} \}.
\end{equation}
Here, we are restricting $\Delta t_{n+1}\in [\Delta t_{\text{min}}, \Delta t_{\text{max}}]$. $F_{\text{max}}>1$ is a bounding scalar factor that prevents $\Delta t_{n+1}$ from increasing too rapidly from $\Delta t_{n}$, and $F_s<1$ is a safety scalar factor to increase the likelihood that $Q^*\leq Q^*_{\text{tol}}$ in the new timestep $\Delta t_{n+1}$. It is necessary to use $F_{\text{max}}$ to restrict the rate of increase of $\Delta t_{n+1}$. Since we use $\vq$ from the current timestep as an approximation of $\vv_{n+1/2}$ in the next timestep, if $\Delta t_{n+1}$ is too large compared to $\Delta t_{n}$, $\vq$ may not be a good approximation even if $Q^*$ is small. This problem does not exist if $\Delta t_{n+1}<\Delta t_n$, and therefore we do not use a bounding factor for the rate of decrease in step size.

Our scheme does not involve rejection of the current timestep $\Delta t_n$, even if in the current timestep $Q^*>Q^*_{\text{tol}}$. Allowing for rejected steps could be considered in future work, although this would increase the complexity of the algorithm, as a copy of the simulation fields at $t_n$ would need to be retained. In practice, the tolerance is only exceeded rarely using the current scheme, usually happening at the onset of plastic deformation. This can be seen in our numerical tests detailed in Sec.~\ref{sec: Numerical results for adaptive globaltimestepping scheme}: as shown in Fig.~\ref{fig:adaptive_scheme_accuracy_plots_A}(a), when $\Bar{s}$ just reached $s_Y$, a peak of $Q^*>Q^*_{\text{tol}}$ of the adaptive timestepping scheme appeared. It was quickly adjusted to be within $Q^*_{\text{tol}}$ for the rest of the simulation. 
To be consistent with the error control of our adaptive global timestepping scheme, during explicit timestepping in \ref{stage1step1.timestepping} in the improved projection method, we control the local timestepping error to be $O(h^2)$ by using the RK2(1) scheme. The RK2(1) scheme and its error control are detailed in \ref{sec: RK21FSAL}.

The adaptive global timestepping procedure automatically resolves evolution time intervals to be finer when there are larger changes to the solution fields. For example, the timesteps are finer when $\Bar{s}$ just reaches $s_Y$ and plastic deformation just starts to happen, or when the amount of plastic deformation increases during some time intervals. On the other hand, when there are not many changes happening in the solution fields (\textit{e.g.}\@ when $\Bar{s}<s_Y$), the simulation takes larger timesteps without undermining the accuracy of solutions. Therefore, the global adaptive timestepping procedure is computationally efficient. It allows us to achieve high accuracy of solutions with significantly fewer timesteps.

\section{Numerical tests}
\label{sec: numerical tests}
We performed numerical tests to verify the second-order temporal accuracy of the improved projection method presented in Sec.~\ref{sec:Second-order projection method for quasi-static elastoplasticity}. We also performed numerical tests to analyze the performance of the adaptive global timestepping procedure presented in Sec.~\ref{sec:Adaptive global timestepping}. Our numerical tests use the 2D plane-strain, simple shear configuration described in Sec.~\ref{sec: set up of numerical tests}. The FEM formulation introduced in Sec.~\ref{sec: FEM projection} is also used to solve the projection steps.

All of our numerical tests use the same simulation setup. The simulation domain is $(x,y)\in [a_x,b_x]\times[a_y,b_y] = [-L,L]\times[-L,L]$, where $L=\SI{1}{\centi\metre}$. Suppose the simulation has grid size $N\times N$, then the grid spacing is $\Delta x=\Delta y=h=2L/N$. A natural timescale of the simulations is $t_s=L/c_{s}=\SI{4.05}{\micro\second}$. Upper and lower wall boundaries are imposed with velocities $\vv_B^{\text{upper}}=(u_B^{\text{upper}},v_B^{\text{upper}})=(U,0)$ moving to the right, and $\vv_B^{\text{lower}}=(u_B^{\text{lower}},v_B^{\text{lower}})=(-U,0)$ moving to the left, respectively, where $U=2 \times 10^{-7} L/t_s=\SI{494}{\micro\metre\per\second}$. The simulation duration is $t= 35 \cdot 10^4 t_s=\SI{1.4175}{\second}$.

The velocity field is initialized at time $t=0 t_s$ to be a smooth gradient field from the lower wall velocity to the upper wall velocity, $\vv_0=(u_0^{i,j},v_0^{i,j})=(-U+j\cdot h\cdot U/L,0)$, where $i, j=1,\ldots,N$, and $i$ denotes the index of the horizontal grid cells, and $j$ denotes the vertical ones. The stress field is initialized as $\mathbf{0}$ everywhere. The initial $\chi$ field is set to be smooth and infinitely differentiable, by using exponential functions. Its initialization details are provided in \ref{appd: numerical test chi init}.

\begin{figure}[!h]
  \centering
  \includegraphics[width=0.8\textwidth]{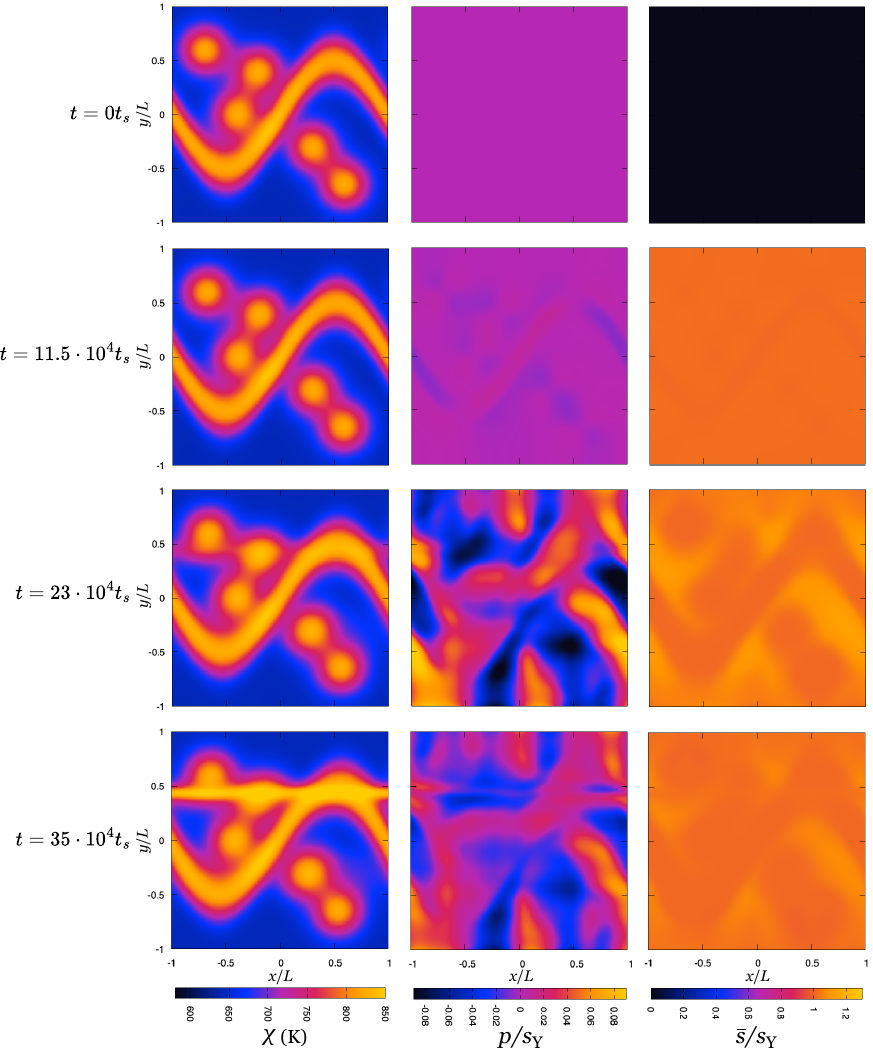}
  \caption{Snapshots of the effective temperature $\chi$, the scaled non-dimensionalize pressure $p/s_Y$ and the scaled non-dimensionalize magnitude of the deviatoric stress $\Bar{s}/s_Y$ in a shearing simulation. The simulation was obtained by using an $N=360$ resolution grid, with constant timestep size $\Delta t=0.5 \cdot 10^4 \cdot h\cdot  t_s/L$. The FEM procedure described in Sec.~\ref{sec: FEM projection} is used for the projection step.
  \label{fig:sim_chi_curves_snapshots}}
\end{figure}

Figure~\ref{fig:sim_chi_curves_snapshots} shows selected time snapshots of the simulation fields, including the effective temperature $\chi$, the pressure $p$, and the magnitude of deviatoric stress $\Bar{s}$. At time $t=0 t_s$, the initial $\chi$ field is designed so that it has different regions of higher and lower $\chi$, and the regions are curved and intersected. Before plastic deformation happens, $\Bar{s}$ increases linearly and evenly across the domain, due to linear elasticity. At around time $t=11.5\cdot 10^4 t_s$, $\Bar{s}$ reached the yield stress $s_Y$ and plastic deformation starts to happen. As seen from the snapshot, the onset of plasticity results in non-linear and uneven changes in the simulation fields. The most obvious one to observe is the $p$ field, where we see positive and negative pressure regions corresponding to material compression and expansion. At time $t=23\cdot 10^4 t_s$, we can see from the $\chi$ field that plastic deformation is propagating from the higher $\chi$ region into the lower $\chi$ region, along the direction of shearing. We also observe growing nonlinear differences in $p$ and $\Bar{s}$, which contributes to more plastic deformation through the plasticity model. At the end of time $t=35\cdot 10^4 t_s$, we can see from the $\chi$ field that a shear band has formed. The formation of the shear band effectively alleviates $p$ and $\Bar{s}$ in the domain. 

\subsection{Numerical results for second-order projection method}
\label{sec: second-order projection-results}
To verify the order of accuracy of the improved projection method described in Sec.~\ref{sec:Second-order projection method for quasi-static elastoplasticity}, we do two sets of tests. In the first set of tests, we use a reduced model and remove the advection and spin terms in the time evolution equations. In this way, we can single out the plasticity terms and verify that they are second-order accurate using the projection method. Therefore, Eqs.~\eqref{eqn:2nd order, intermediate stress, stage 1} \& \eqref{eqn:2nd order, chi, stage 1} become
\begin{equation}
\label{eqn:2nd order, intermediate stress, only plas}
(\bsig)_t=\bC:\bD_{\vq}-\bC:\bD^{\text{pl}}(\bsig, \chi)
\end{equation}
and
\begin{equation}
\label{eqn:2nd order, chi, only plas}
(\chi)_t=F(\bsig, \chi).
\end{equation}
In the second set of tests, we use the full model and solve the system given in Eqs.~\eqref{eqn:2nd order, intermediate stress, stage 1} \& \eqref{eqn:2nd order, chi, stage 1}, and verify that the scheme is second-order accurate with all terms added in.

For each set of tests, we first simulate on a $N=2520$ grid to calculate reference solutions $\vv_r$, $\bsig_r$ and $\chi_r$. We then run simulations on grids with sizes $N \in\{280, 360, 504\}$. We set $\Delta t=0.5 \cdot 10^4 \cdot h\cdot  t_s/L$ for each simulation case. The RK2(1) scheme used to explicitly timestep \ref{stage1step1.timestepping} of the improved projection method uses $A_{\text{tol}}=\max (F_{\text{RK2(1)}} \Delta t^2, 10^{-16})$ and $R_{\text{tol}}=\max (F_{\text{RK2(1)}} \Delta t^2, 10^{-17})$, where $F_{\text{RK2(1)}}=1e^{-7}$. The RK2(1) scheme, therefore, bounds the local sub-stepping error to be $O(\Delta t^2)$, consistent with the second-order temporal error of $O(\Delta t^2)$ of the projection method.

To calculate the simulation errors, we define a norm on simulation fields $\vec{f}$,
\begin{equation}
\label{eqn:2nd order, error norm}
\lVert \vec{f}(t) \rVert = \sqrt{\frac{1}{A}\int_{a_x}^{b_x}\int_{a_y}^{b_y} |\vec{f}(\vx,t)|^2 \,dx \, dy}
\end{equation}
where $A=(b_x-a_x)(b_y-a_y)$ is the area of the domain, and $|\cdot|$ is taken to be the Euclidean norm for vectors, absolute value for scalars, and the Frobenius norm for tensors. The integral can be computed via the trapezoid rule over the domain. For each simulation case of grid size $N \in\{280, 360, 504\}$, we use Eq.~\eqref{eqn:2nd order, error norm} to evaluate the norms for the error fields at the end of the simulation time, $\vv_i-\vv_r$, $\bsig_i-\bsig_r$ and $\chi_i-\chi_r$. The error norms are non-dimensionalized by dividing $U$, $s_Y$ and $\chi_{\infty}$, respectively.

\begin{figure}[!h]
  \centering
  \includegraphics[width=0.5\textwidth]{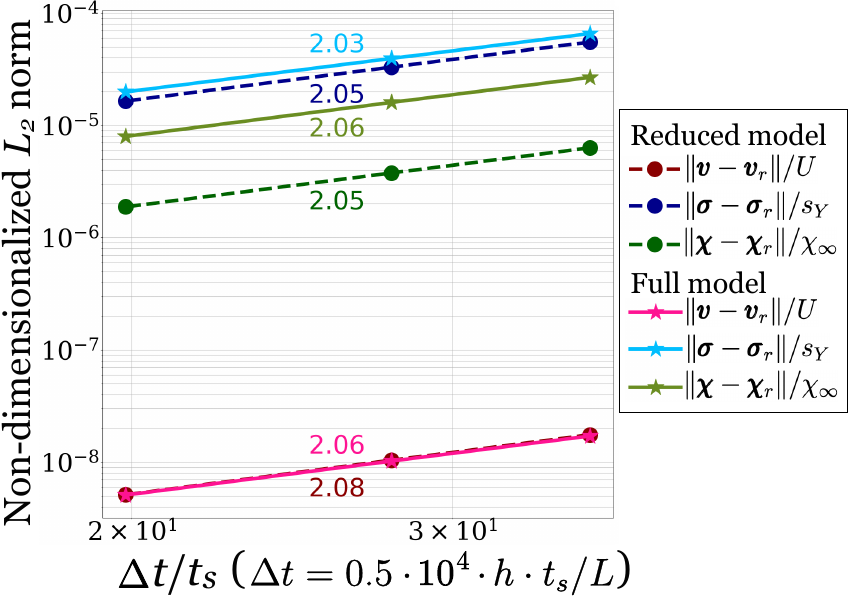}
  \caption{The accuracy plot for the reduced model and the full model. In the plot, the errors for the three fields, $E_{\vv}$, $E_{\bsig}$ and $E_{\chi}$, are non-dimensionalized by the scaling $E_{\vv}/U$, $E_{\bsig}/s_Y$ and $E_{\chi}/\chi_{\infty}$, respectively. The slopes of the log--log plot are calculated using linear regression of the log--log solution accuracy values from grids of sizes $N=280, 360, 504$. The slope values are labeled in the plot, and we see that all fields reach full second-order accuracy in their solution fields for both the reduced and the full models.\label{fig:sim_chi_init_final_accuracy}}
\end{figure}

The accuracy of the solutions is shown in Fig.~\ref{fig:sim_chi_init_final_accuracy}. The orders of accuracy are annotated in the plot. They are the slopes of the log--log plot, calculated using linear regression of the log--log solution accuracy values from grids of sizes $N=280, 360, 504$. We see that the reduced model has second-order temporal accuracy for all fields, therefore verifying the full second-order accuracy of the projection method on the plasticity terms. For the full model, the errors of $\bsig$ and $\chi$ have shifted up, and the difference is especially prominent for $\chi$. From testing, we found that the advection terms have larger magnitudes of errors, therefore dominating the errors for $\chi$ in the full model. However, we verified in the first set of tests that the plasticity terms achieve second-order accuracy. Here, with the full model, we have further verified that all terms in the PDE system reach second-order accuracy using the projection scheme.

\subsection{Numerical results for adaptive global timestepping scheme}
\label{sec: Numerical results for adaptive globaltimestepping scheme}
We use the adaptive global timestepping scheme as described in Sec.~\ref{sec:Adaptive global timestepping} and simulate the full model on grids with sizes $N \in\{280, 360, 504\}$. The default values used for control parameters of the scheme are listed in Table~\ref{tbl:Adaptive global timestepping parameters}. The RK2(1) scheme used to explicitly timestep \ref{stage1step1.timestepping} of the improved projection method for the adaptive global timestepping scheme uses $A_{\text{tol}}=\max (F_{\text{RK2(1)}} h^2/L^2, 10^{-16})$ and $R_{\text{tol}}=\max (F_{\text{RK2(1)}} h^2/L^2, 10^{-17})$, where $F_{\text{RK2(1)}}=0.1$. The RK2(1) scheme, therefore, bounds the local sub-stepping error to be $O(h^2)$, comparable to the spatial discretization error of $O(h^2)$. We use the same simulation setup described in Sec.~\ref{sec: second-order projection-results}. The constant timestepping simulation on the $N=2520$ grid in Sec.~\ref{sec: second-order projection-results} is used as a reference solution. We compare the adaptive global timestepping simulations with the constant timestepping simulations from Sec.~\ref{sec: second-order projection-results}.

\begin{table}[h!]
\centering
\begin{tabular}{|p{20mm}|p{20mm}|p{20mm}|p{20mm}|p{20mm}|p{20mm}|}
\hline
    $F_Q$ & $0.001$ & $F_{\text{max}}$ & $2$ & $F_s$ & $0.9$
\\ \hline
    $\Delta t_{\text{min}}$ & $10^{-6} t_s$ & $\Delta t_{\text{max}}$ & $10^3 t_s$ & $\Delta t_{\text{init}}$ & $10 t_s$
\\ \hline

\end{tabular}
\caption{Adaptive global timestepping scheme parameters used \label{tbl:Adaptive global timestepping parameters}}
\end{table}

\begin{figure}
  \centering
  \includegraphics[width=0.9\textwidth]{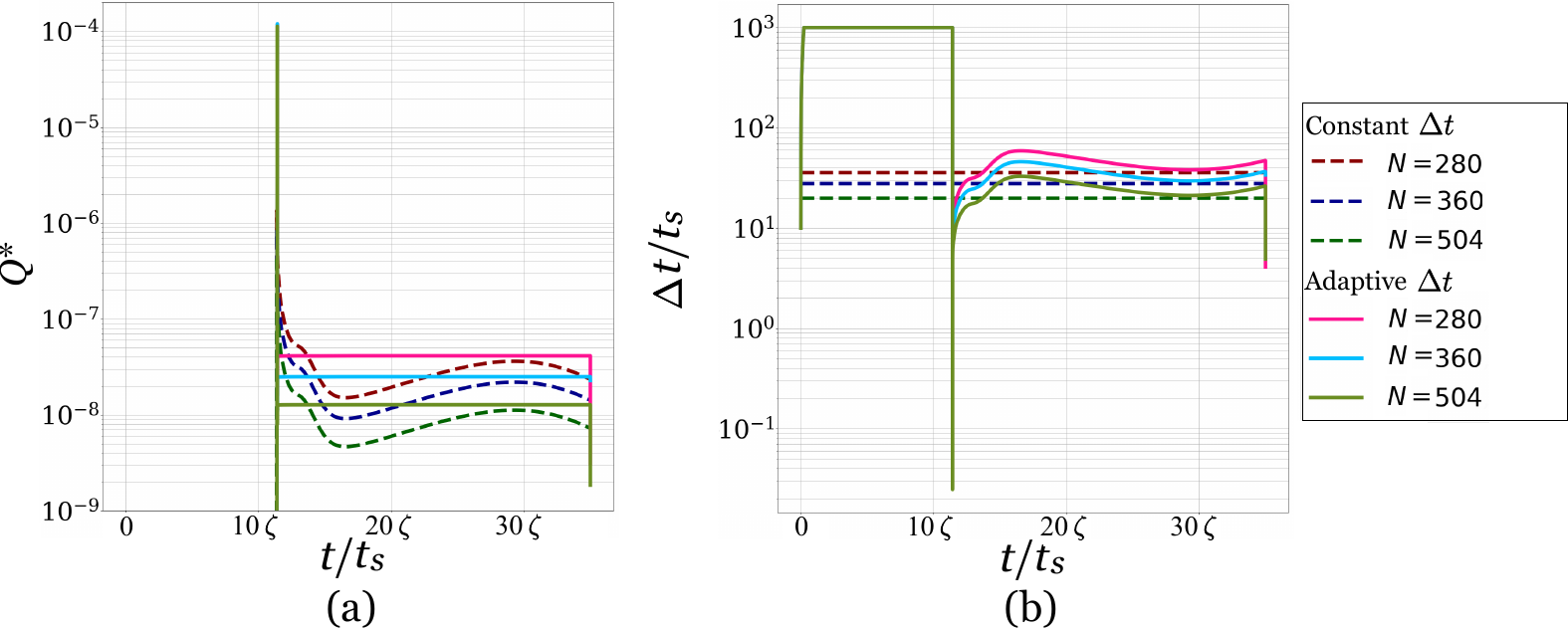}
  \caption{(a) Comparison of the evolution of $Q^{*}$ versus time $t/t_s$, of the constant timestepping scheme $\Delta t=0.5 \cdot 10^4 \cdot h\cdot t_s/L$, and the adaptive timestepping scheme using $Q^{*}_{\text{tol}}=0.001 h^2/L^2$; For the adaptive timestepping scheme, $Q^{*}$ exceeded $Q^{*}_{\text{tol}}$ only rarely, at the onset of plastic deformation around time $t=11.5\cdot 10^4 t_s$. The scheme then quickly adjusted $\Delta t$ and $Q^{*}$ remained within $Q^{*}_{\text{tol}}$ for the rest of the simulation. (b) Comparison of the $\Delta t$ step size over time for the two schemes.~\label{fig:adaptive_scheme_accuracy_plots_A}}
\end{figure}

Figure~\ref{fig:adaptive_scheme_accuracy_plots_A}(a) shows the evolution of $Q^*$ over time. In the beginning, when $\Bar{s}<s_Y$ and there was little change in the simulation fields, both constant and adaptive timestepping simulations had very small $Q^*\approx0$. Then a sudden peak in $Q^*$ occured at around time $t=11.5 \cdot 10^4 t_s$, corresponding to the onset of plastic deformation when $\Bar{s}$ just surpassed $s_Y$. Afterward, for the constant timestepping cases, $Q^*$ decreased and varied over time. For the adaptive timestepping cases, since we do not have the rejection procedure on timesteps, the initial peaks of $Q^*$ surpassed the tolerances $Q^*_{\text{tol}}$. However, for each case, $Q^*$ rapidly decreased and maintained at the level bounded by $Q^*_{\text{tol}}$. Figure~\ref{fig:adaptive_scheme_accuracy_plots_A}(b) shows the timestep sizes for the simulations over time. The constant timestepping cases had $\Delta t=0.5 \cdot 10^4 \cdot h\cdot t_s/L$ throughout the simulations. For the adaptive timestepping cases, when $\Bar{s}<s_Y$, $\Delta t=\Delta t_{\max}$. When $\Bar{s}\geq s_Y$ and plastic deformation occurred, $\Delta t$ was adjusted accordingly to control $Q^*$ based on $Q^*_{\text{tol}}$.

\begin{figure}
  \centering
  \includegraphics[width=0.9\textwidth]{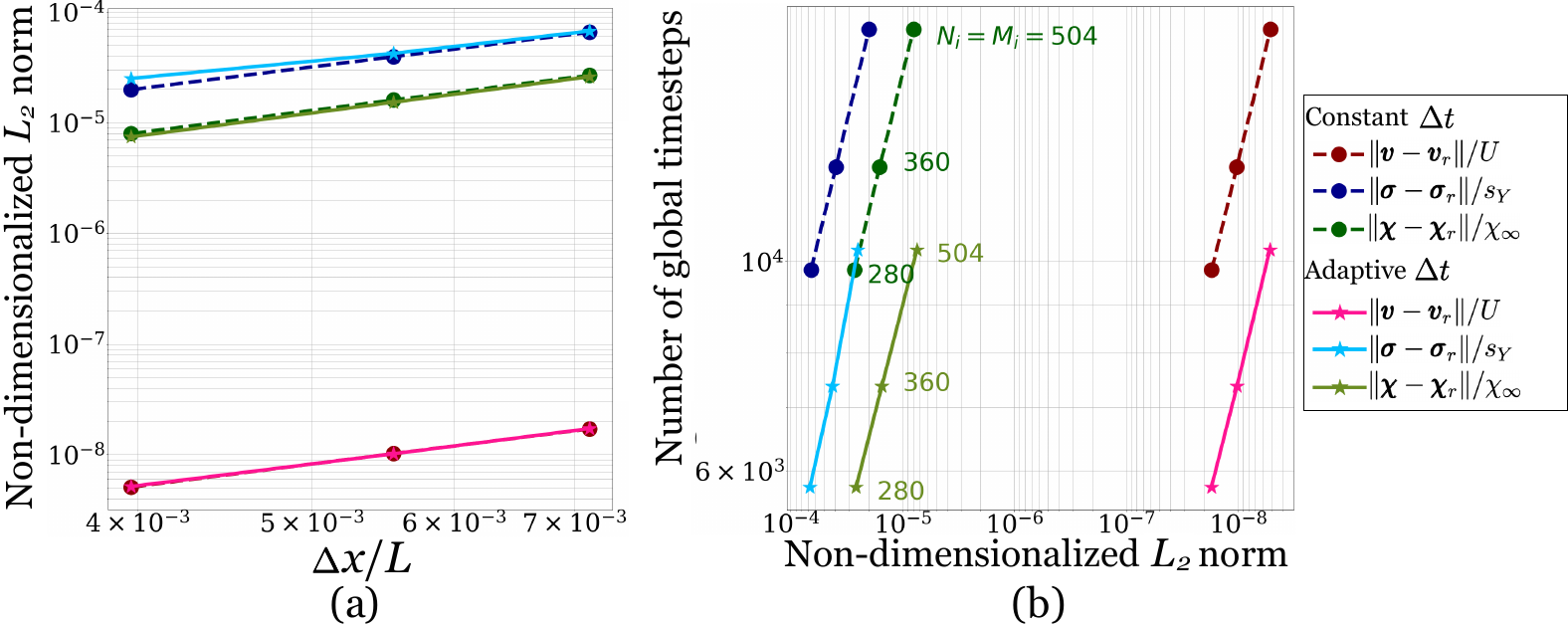}
  \caption{(a) Comparison of accuracy of solution fields. Both the constant and the adaptive timestepping schemes reached the same level of solution accuracy; (b) Comparison of the number of timesteps taken for the two schemes to reach the same level of solution accuracy. The adaptive timestepping scheme took significantly fewer timesteps. ~\label{fig:adaptive_scheme_accuracy_plots_B}}
\end{figure}

Figure~\ref{fig:adaptive_scheme_accuracy_plots_B}(a) compares the field accuracy of the adaptive and constant timestepping simulations for each grid size case. We see that the two schemes achieve almost identical accuracy in solutions. Figure~\ref{fig:adaptive_scheme_accuracy_plots_B}(b) provides the accuracy of each simulation versus the number of global timesteps it takes. To achieve the same level of accuracy, the adaptive timestepping simulations use only about half the number of timesteps for each grid case compared to the constant timestepping simulations, resulting in a significant computational saving.

\section{Conclusion}
\label{sec:conclusion}
Motivated by accurate and efficient simulations of the mechanical deformation of BMGs, we made several numerical improvements in modeling QS elastoplastic materials. Firstly, drawing inspiration from incompressible fluid dynamics, we developed a second-order projection method for QS elastoplasticity. This is an improvement from the original projection method that is first-order accurate in time. Furthermore, we implemented an FEM formulation of the projection step to solve the elliptic PDEs in the method, which provides numerical benefits in solving the linear system and in implementing boundary conditions. Lastly, we created an adaptive global timestepping procedure, which improves computational efficiency and allows us to achieve highly accurate solutions in significantly fewer timesteps. We performed two-dimensional numerical tests to evaluate each of the improvements, using an example physical model of a BMG based on an STZ plasticity model. Though we focus on two-dimensional simulations of BMGs in this paper, the numerical methods and improvements can be applied to any elastoplastic materials and extended to three dimensions.

In the future, we plan to investigate higher-order numerical formulations of the projection method for solving QS elastoplasticity. For a spatial discretization, we can use the discontinuous Galerkin (DG) method~\cite{Shu2009DG, saye2017I,saye2017II,hesthaven2007nodal}. Using piecewise-polynomial functions of degree $s$, defined on each element in the simulation mesh, we can obtain high-order spatial accuracy depending on the choice of $s$.

For high-order temporal accuracy, we draw inspiration from work by Minion and Saye~\cite{MINION2018} on developing the spectral deferred pressure correction method (SPDC), a numerical scheme based on the fluid projection method that can reach an arbitrarily high order of temporal accuracy in solving the incompressible Navier--Stokes equations. SPDC decomposes each timestep $\Delta t = t_{n+1} - t_n$ into substeps. The substeps are chosen as Gauss--Lobatto quadrature nodes, and are used to integrate in time using quadrature rules. The time-stepping is also iterative. In each timestepping from $t_n$ to $t_{n+1}$, iterations of the substepping are performed. Furthermore, at the end of each substep in each iteration, the projection step is used to project the velocity field to maintain the incompressibility constraint. For the QS elastoplasticity system, we can try transferring the SPDC scheme over. We can perform the projection scheme over a timestep with substeps and iterations, and investigate the effect on numerical accuracy.

\section*{Acknowledgements}
This work was supported by the National Science Foundation under Grant No.~DMS-1753203. C.~H.~R.~was partially supported by the Applied Mathematics Program of the U.S.~DOE Office of Science Advanced Scientific Computing Research under Contract No.~DE-AC02-05CH11231.

\appendix

\section{STZ plasticity model}
\label{appendix: plasticity model}
To describe plasticity of the material, that is, $\bD^{\text{pl}}$ in Eq.~\eqref{eqn:linear constitutive eqn}, we use a plasticity model based on the athermal form of the shear transformation zone (STZ) theory of amorphous plasticity~\cite{Bouchbinder2007STZ,Langer2008STZ}. STZs are localized regions susceptible to configurational changes by external shear~\cite{Bouchbinder2007STZ}. In thermal theories~\cite{Kamrin2014thermo}, a configurational subsystem governs the rearrangements that occur at STZs, and a kinetic/vibrational subsystem governs the thermal vibrations of atoms in their cage of nearest neighbors. Furthermore, the two subsystems are coupled. In the athermal form of the STZ theory, thermal fluctuations of the atomic configuration are neglected, and molecular rearrangements are only driven by external mechanical forces.
It is assumed that a population of STZs exist in an otherwise elastic material. Under external shear, if the local stress surpasses the material yield stress $s_Y$, localized irreversible molecular rearrangements occur sporadically at the STZ sites. These molecular rearrangements contribute to small increments in strain, and their accumulation leads to macroscopic plastic deformation.

An effective disorder temperature~\cite{Bouchbinder2011STZ,Bouchbinder2008Chi,Loi2008Chi,Cugliandolo2011Chi}, $\chi$, is used to describe the density of STZs in the material. $\chi$ is measured in Kelvin and follows a Boltzmann distribution in space. $\chi$ can be obtained as the derivative of a configurational energy with respect to a configurational entropy~\cite{Falk2011Chi}. $\chi$ is different from the thermodynamic temperature $T$, but it serves similar purposes for the configurational subsystem as $T$ does for the kinetic/vibrational subsystem.

\begin{table}
\centering
\begin{tabular}{|p{80mm}|p{30mm}|}
\hline
    Molecular vibration timescale $\tau_0$&$10^{-13}$ \si{s}
\\ \hline
    Typical local strain $\epsilon_0$&$0.3$
\\ \hline
    Effective heat capacity $c_0$&$0.4$
\\ \hline
    Typical activation barrier $\Delta/k_B$&$8000$ \si{K}
\\ \hline
    Typical activation volume $\zeta$&$300$ \AA$^3$
\\ \hline
    Thermodynamic bath temperature $T$&$400$ \si{K}
\\ \hline
    Steady state effective temperature $\chi_{\infty}$&$900$ \si{K}
\\ \hline
    STZ formation energy $e_Z/k_B$&$21000$ \si{K}
\\ \hline

\end{tabular}
\caption{Plastic deformation parameters for the STZ plasticity model used. The Boltzmann constant $k_B=1.3806488\times 10^{-23}$ \si{JK^{-1}} is used to convert energy values to temperatures. \label{tbl:plasticity parameters}}
\end{table}

The plastic deformation tensor $\bD^\text{pl}$ is proportional to the deviatoric stress tensor, and is given in Eq.~\eqref{eqn:Dpl tensor equation}, $\bD^\text{pl}=\frac{\bsig_0}{\Bar{s}}D^\text{pl}$, where $D^\text{pl}$ is a scalar function of $\chi$ and $\Bar{s}$. When $\Bar{s}<s_Y$, $D^\text{pl}=0$. When $\Bar{s}\geq s_Y$, plastic deformation is given by

\begin{equation}
\label{appdix eqn:Dpl calculation, main}
D^{\text{pl}}(\bsig_0,T,\chi)=\frac{\Lambda(\chi)\mathcal{C}(\Bar{s},T)}{\tau_0}(1-\frac{s_Y}{\Bar{s}}),
\end{equation}
where $\tau_0$ is a molecular vibration timescale, $\mathcal{C}(\Bar{s},T)$is the STZ transition rate and $\Lambda(\chi)$ is the density of STZs in terms of effective temperature $\chi$. The relevant parameters for the plasticity deformation calculation are given in Table~\ref{tbl:plasticity parameters}, used throughout this paper. Using these parameters, the functions in Eq.~\eqref{appdix eqn:Dpl calculation, main} are
\begin{equation}
\label{appdix eqn:Dpl calculation, Lambda function}
\Lambda(\chi)=e^{-e_z/(k_B\chi)},
\end{equation}
and
\begin{equation}
\label{appdix eqn:Dpl calculation, C function}
\mathcal{C}(\Bar{s},T)=e^{-\Delta/(k_B T)}\cosh{\frac{\zeta\epsilon_0\Bar{s}}{k_B T}}.
\end{equation}

The effective temperature $\chi$ responds to the plastic deformation as given in Eq.~\eqref{eqn:Chi evolution}. Under externally applied mechanical work, STZs are created and annihilated proportionally, and $\bD^{\text{pl}}:\bsig_0$ describes the energy dissipation rate of this process. $\chi$ therefore evolves like a heat equation, corresponding to the first law of thermodynamics for the configurational subsystem~\cite{Bouchbinder2007STZ}. Furthermore, according to Eq.~\eqref{eqn:Chi evolution}, an increase in plastic deformation $\bD^\text{pl}$ increases $\chi$, until it saturates at $\chi_{\infty}$. On the other hand, from Eq.~\eqref{appdix eqn:Dpl calculation, main} \& \ref{appdix eqn:Dpl calculation, Lambda function}, an increase in $\chi$ increases $D^\text{pl}$, which then increases $\bD^{\text{pl}}$ via Eq.~\eqref{eqn:Chi evolution}. The mutual feedback of the plasticity model typically leads to shear banding~\cite{Manning2007STZ,Manning2009STZ}.

\begin{table}
  \[
    \renewcommand\arraystretch{1.2}
    \begin{array}
    {c|ccc}
    k_1:\quad 0\\
    k_2:\quad \frac{2}{3} & \frac{2}{3}\\
    k_3:\quad 1 & \frac{1}{4}&\frac{3}{4} \\
    \hline
    2^{\text{nd}} \text{ order: } \hat{\vy}_{1}& \frac{1}{4} &\frac{3}{4} \\
    1^{\text{st}} \text{ order: } \vy_{1}& \frac{1}{3} &\frac{1}{3} &\frac{1}{3}
    \end{array}
    \]
  \caption{The Butcher tableau for an adaptive RK2(1) method}
  \label{tab:RK21FSAL}
\end{table}

\section{An explicit RK 2(1) FSAL adaptive timestepping scheme}
\label{sec: RK21FSAL}
To solve the coupled ODE system of Eqs.~\eqref{eqn:2nd order, intermediate stress, stage 1} \& \eqref{eqn:2nd order, chi, stage 1}, and achieve second order temporal accuracy in the timestepping, we use an explicit RK2(1) FSAL adaptive timestepping scheme~\cite{ralston1965,runge1895RKmethod,kutta1901RKmethod}. It is also capable of handling the numerical challenges given by the plasticity STZ model discussed in Sec.~\ref{sec: plasticity model and its numerical challenges}.
The associated Butcher tableau is given in Table~\ref{tab:RK21FSAL}. Consider starting from $\vy_0$ and taking a step of size $\Delta t$. The second-order solution $\hat{\vy}_{1}$ is used to integrate the ODE, and uses the Ralston method using the first two stages, $k_1$ and $k_2$. The first-order solution $\vy_{1}$ is used for step size selection, and also makes use of the third stage $k_3$. The method has the FSAL property so that $k_3$ can be reused as $k_1$ at the next timestep for computational efficiency~\cite{hairer2008solving}.
Since $\hat{\vy_1}$ is more accurate, $\hat{\vy}_{1}-\vy_{1}$ is used as an error estimate of $\vy_1$ compared to the true solution. We aim to satisfy
\begin{equation}
\label{eqn: Rk21 criteria}
|y_{1,i}-\hat{y}_{1,i}|<s_i, \quad s_i=A_{\text{tol},i}+R_{\text{tol},i}\max(|y_{0,i}|,|\hat{y}_{0,i}|),
\end{equation}
where $i$ denotes the components of the solution, and $A_{\text{tol},i}$ and $R_{\text{tol},i}$ are the absolute and relative error tolerances, respectively. If there are $n$ degrees of freedom of the solution, the scaled measure of error is given by
\begin{equation}
\label{eqn: Rk21 scaled error}
E=\sqrt{\frac{1}{n}\sum_{i=1}^n\left( \frac{y_{1,i}-\hat{y}_{1,i}}{s_i} \right)^2}.
\end{equation}
Since $\vy_1$ is first-order accurate, the local error over a single timestep will scale quadratically, according to $E\approx c\cdot (\Delta \hat{t})^{2}$. We require $E<1$ for an acceptable time substep. Therefore, given an estimate of $E$ from the previous step, the predicted optimal step size is
\begin{equation}
\label{eqn: Rk21 optimal h}
\Delta \hat{t}_{\text{opt}}=\Delta \hat{t}E^{-1/2},
\end{equation}
since this should result in new $E$ close to 1. We further scale $\Delta \hat{t}_{\text{opt}}$ with a safety factor $f_s<1$, so that the timestep will be accepted with high probability. We also bound $\Delta \hat{t}_{\text{opt}}$ with factors $f_{\min}$ and $f_{\max}$, so that it does not change by a large amount from the previous time step. Therefore, we can calculate $\Delta \hat{t}_{\text{new}}$ as
\begin{equation}
\label{eqn: Rk21 optimal h new}
\Delta \hat{t}_{\text{new}}=\Delta \hat{t}\cdot\min \{f_{\max}, \max\{f_{\min}, f_s \cdot E^{-1/2}\} \}.
\end{equation}
If $E<1$, the current time step $\Delta \hat{t}$ is accepted, and a new time step $\Delta \hat{t}_{\text{new}}$ is tried. Otherwise, the current $\Delta \hat{t}$ is rejected, and we solve for the solution again from $\vy_0$ with a new time step size $\Delta \hat{t}_{\text{new}}$. From the formula given in Eq.~\eqref{eqn: Rk21 optimal h new}, $\Delta \hat{t}_{\text{new}}<\Delta \hat{t}$ in the case of a rejection. The $\Delta \hat{t}_{\text{new}}$ is tried following the same error calculation and criteria, and the process repeats until we find a satisfying $\Delta \hat{t}_{\text{new}}$. We chose parameters of $f_{\min}=\frac13$, $f_{\max}=3$, and $f_s=0.9$.

The choices of $A_{\text{tol}}$ and $R_{\text{tol}}$ depend on the problem. In Sec.~\ref{sec:Second-order projection method for quasi-static elastoplasticity}, where we solve for $\bsig_*$ and $\chi_{n+1}$ in \ref{stage1step1.timestepping} for the second-order projection method, we use $A_{\text{tol}}=\max ( F_{\text{RK2(1)}} \Delta t^2, 10^{-16})$ and $R_{\text{tol}}=\max (F_{\text{RK2(1)}} \Delta t^2, 10^{-17})$, where $F_{\text{RK2(1)}}$ is a scalar factor and $\Delta t$ is the fixed full timestep size of the method. $A_{\text{tol}}$ and $R_{\text{tol}}$ are set in this way, so that they bound the local timestepping error to be $O(\Delta t^2)$, consistent with the second-order in time projection method.

For the problem in Sec.~\ref{sec:Adaptive global timestepping criteria and procedure}, where we develop an adaptive global timestepping scheme, there is no longer a fixed global timestep $\Delta t$. Instead, we use $A_{\text{tol}}=\max (F_{\text{RK2(1)}} h^2, 10^{-16})$ and $R_{\text{tol}}=\max (F_{\text{RK2(1)}} h^2, 10^{-17})$ for the RK2(1) scheme, where $F_{\text{RK2(1)}}$ is a scalar factor. They are set in this way to bound the local sub-stepping error to be $O(h^2)$, comparable to the spatial discretization error of $O(h^2)$. Since the overall numerical accuracy is bounded by the second-order spatial discretization, there is no need to make $A_{\text{tol}}$ and $R_{\text{tol}}$ to be of higher order accuracy.

\section{Initialization of $\chi$ field}
\label{appd: numerical test chi init}
This appendix describes the initialization of the $\chi$ field for the numerical tests in Sec.~\ref{sec: numerical tests}. 
We first chose a base temperature $T_{\text{base}}$ and a temperature change lengthscale $\Delta T$. We then define
\begin{align}
  \chi_0&=\frac{T_{\text{base}}}{e_z/k_B}, \\
  \Delta\chi&=\frac{\Delta T}{e_z/k_B},
\end{align}
where $e_z/k_B$ is the STZ formation energy given in Table~\ref{tbl:plasticity parameters}, and the values of $T_{\text{base}}$ and $\Delta T$ are given in Table~\ref{tbl:chi init numerical test}. At each grid cell of the simulation, let $\vec{x}_m=(x_m,y_m)$ denote the midpoint of the cell. The $\chi$ value of the cell is calculated as
\begin{equation}
\chi(\vec{x}_m) = \chi_0+\chi_1 \sum_{i=1}^{6} e^{-20\cdot R_i(\vec{x}_m)},
\end{equation}
where $R_1(\vec{x}_m)=Y_{r,1}(\vec{x}_m)^2$, and $R_i(\vec{x}_m)=X_{r,i}(\vec{x}_m)+(y_m+s_i)$ for $i=2, \cdots ,6$. The related parameters ($X_{r,i}(\vec{x}_m)$, $Y_{r,i}(\vec{x}_m)$, $s_i$) and their chosen values are provided in Table~\ref{tbl:chi init numerical test2}.

\begin{table}[h!]
\centering
\begin{tabular}{|p{70mm}|p{50mm}|}
\hline
    A base temperature $T_{\text{base}}$&$640$ \si{K}
\\ \hline
    A temperature change lengthscale $\Delta T$&$180$ \si{K}
\\ \hline
\end{tabular}
\caption{Parameters used in the initialization of $\chi$ field for the numerical tests in Sec.~\ref{sec: numerical tests}. \label{tbl:chi init numerical test}}
\end{table}

\begin{table}[h!]
\centering
\begin{tabular}{|p{6mm}|p{20mm}|p{40mm}|p{15mm}|p{50mm}|}
\hline
    $i$&$X_{r,i}(\vec{x}_m)$&$Y_{r,i}(\vec{x}_m)$&$s_i$&Note on $X_{r,i}(\vec{x}_m)$ calculation
\\ \hline
    $1$& -- & $0.5\cdot \sin{(\pi \cdot x_m)} - y_m$ & -- & --
\\ \hline
    
\multicolumn{1}{|l|}{$2$} & \multicolumn{1}{|l|}{$x_m-0.6$} & \multicolumn{1}{|l|}{--} & \multicolumn{1}{|l|}{$0.65$} & \multirow{5}{50mm}{Due to periodicity in the $x$-direction, if $X_{r,i}(\vec{x}_m)<a_x$, $X_{r,i}(\vec{x}_m)$ is replaced by $X_{r,i}(\vec{x}_m)+2L$; if $X_{r,i}(\vec{x}_m)>b_x$, $X_{r,i}(\vec{x}_m)$ is replaced by $X_{r,i}(\vec{x}_m)-2L$.} \\ \cline{1-4}

\multicolumn{1}{|l|}{$3$} & \multicolumn{1}{|l|}{$x_m-0.3$} & \multicolumn{1}{|l|}{--} & \multicolumn{1}{|l|}{$0.3$} & \\ \cline{1-4}

\multicolumn{1}{|l|}{$4$} & \multicolumn{1}{|l|}{$x_m+0.2$} & \multicolumn{1}{|l|}{--} & \multicolumn{1}{|l|}{$-0.4$} & \\ \cline{1-4}

\multicolumn{1}{|l|}{$5$} & \multicolumn{1}{|l|}{$x_m+0.4$} & \multicolumn{1}{|l|}{--} & \multicolumn{1}{|l|}{$0$} & \\ \cline{1-4}

\multicolumn{1}{|l|}{$6$} & \multicolumn{1}{|l|}{$x_m+0.7$} & \multicolumn{1}{|l|}{--} & \multicolumn{1}{|l|}{$-0.6$} & \\ \cline{1-5}

\end{tabular}
\caption{Parameters used in the initialization of $\chi$ field for the numerical tests in Sec.~\ref{sec: numerical tests}. \label{tbl:chi init numerical test2}}
\end{table}

\color{black}

\bibliography{refs}

\clearpage

\setcounter{section}{0}
\setcounter{equation}{0}
\setcounter{figure}{0}
\setcounter{table}{0}
\setcounter{page}{1}
\makeatletter
\renewcommand{\theequation}{S.\arabic{equation}}
\renewcommand{\thesection}{S.\arabic{section}}    
\renewcommand{\thefigure}{S.\arabic{figure}}
\renewcommand{\thetable}{S.\arabic{table}}

\begin{center}
    \large{Electronic supplementary material for\\
``Numerical methods and improvements for simulating quasi-static elastoplastic materials'': \\Finite element (FEM) formulation of the projection step} \\
    \vspace{\baselineskip}
    \normalsize Jiayin~Lu\textsuperscript{a,b}, Chris H.~Rycroft\textsuperscript{b,c}\\
    \textsuperscript{a}\textit{Department of Mathematics, University of California, Los Angeles, Los Angeles, CA 90095, USA} \\
    \textsuperscript{b}\textit{Department of Mathematics, University of Wisconsin--Madison, Madison, WI 53706, USA} \\
    \textsuperscript{c}\textit{Mathematics Group, Lawrence Berkeley Laboratory, Berkeley, CA 94720, USA}
\end{center}

\section{Introduction}
In the main paper, we derived a finite element method (FEM)~\cites{donea_fluid_fem,claes_fem} formulation to solve the projection step for both the original first-order projection method and the improved second-order projection method. In Sec.~\ref{sec:Derivation of the FEM projection step}, we provided a weak form formulation of the FEM procedure for the original projection method, where we solve for $\vv_{n+1}$ using Eq.~\eqref{eqn:projection:solving v at time n+1}. The FEM velocity boundary condition uses $\vv^B=(u^B,v^B)$, which is the Dirichlet boundary condition on the upper and lower walls of the 2D simple shear simulation, as described in Sec.~\ref{sec: set up of numerical tests}.

The FEM formulation gives rise to a linear system to solve, $\bA \vw=\vb$. Its component form is derived in detail in \ref{appendix:FEM linear system details}. For the second-order projection method, where we aim to solve for the correction velocity $\bPhi_*$ via Eq.~\eqref{eqn:2nd order, projection, div stress *, stage 1}, the FEM procedure is similar. The only difference, as mentioned in Sec.~\ref{sec: FEM projection}, is that the FEM velocity boundary condition becomes $\bPhi_B=\mathbf{0}$. The FEM linear system associated with the second-order projection method in component form is discussed in Sec.~\ref{appendix: FEM second-order projection}.

\section{FEM projection linear system derivation: Original projection method}
\label{appendix:FEM linear system details}
We discretize Eq.~\eqref{eqn:fem projection, main, rearrange} to obtain a linear system $\bA\vw=\vb$ of $2\mathcal{I}$ equations, to solve for the $2\mathcal{I}$ unknown inner node velocity weights, $\{w_{i,k}^{I}\}_{i = 1, \ldots ,\mathcal{I}; k=1,2}$. This is done by varying the test function in each dimension $k=1,2$ to be $\{\psi_{i}\}_{i = 1, \ldots,\mathcal{I}}$. Here we detail the derivation of the component form of the linear system.

\subsection{Nodal basis functions using the bilinear element}
\label{appendix: fem: nodal basis}
\begin{figure}
    \centering
    \includegraphics[width=1\textwidth]{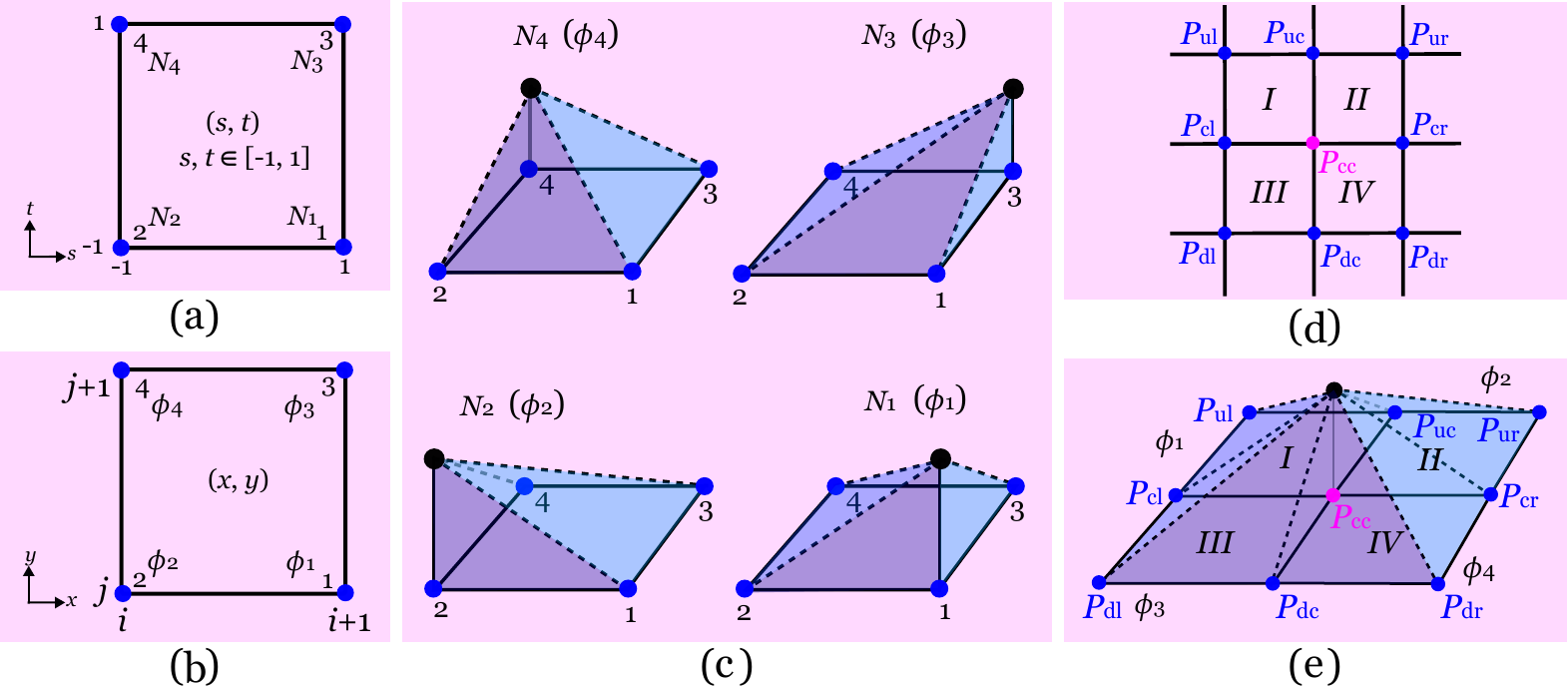}
    \caption{(a) The $(s,t)$ coordinate space, where $s,t\in[-1,1]$. The functions $N_q,q=1,2,3,4$ are defined for the four corner nodes of the domain. (b) The $(x,y)$ coordinate space corresponding to a grid cell $(i,j)$ in our simulation domain. The functions $\phi_q$ for $q=1,2,3,4$ are defined for the four corner nodes of the grid cell. (c) The shapes of the $N_q$ and $\phi_q$ functions defined for the corner nodes $q=1,2,3,4$ in their corresponding domains. The functions are linear in their rectangular domains, and have value $1$ at corner $q$, and $0$ at other corners. (d) An inner node $P_{\text{cc}}$, its neighboring nodes, and its four neighboring grid cell regions. (e) A bilinear element in one dimension, $\psi_{\text{cc}}$, for an inner node $P_{\text{cc}}$. In each of the surrounding regions, $I,II,III,IV$ of $P_{\text{cc}}$, $\psi_{\text{cc}}$ is composed of the corresponding function, $\phi_1,\phi_2,\phi_3,\phi_4$ in the region.}
    \label{fig:FEM_appd_plot}
\end{figure}

First, we introduce four linear functions, $N_q(s,t)$, $q=1,\ldots,4$, defined on $s,t\in[-1,1]$. The four functions and their partial derivatives are given in Table~\ref{tbl:appd, N_i and derivatices}. As shown in Fig.~\ref{fig:FEM_appd_plot}(a)(c), for the four corners labeled $q=1,\ldots,4$ in the $(s,t)$ domain, $N_q$ has value $1$ at corner $q$, and $0$ at the other corners.

\begin{table}
\centering
\begin{tabular}{|p{50mm}|p{30mm}|p{30mm}|}
\hline
    $N_q(s,t), s,t\in[-1,1]$ & $\frac{\p N_i}{\p s}$ & $\frac{\p N_i}{\p t}$
\\ \hline \hline
    $N_1=\frac{1}{4}(1+s)(1-t)$ & $\frac{1-t}{4}$ & $\frac{-s-1}{4}$
\\ \hline
    $N_2=\frac{1}{4}(1-s)(1-t)$ & $\frac{t-1}{4}$ & $\frac{s-1}{4}$
\\ \hline
    $N_3=\frac{1}{4}(1+s)(1+t)$ & $\frac{t+1}{4}$ & $\frac{s+1}{4}$
\\ \hline
    $N_4=\frac{1}{4}(1-s)(1+t)$ & $\frac{-t-1}{4}$ & $\frac{1-s}{4}$
\\ \hline

\end{tabular}
\caption{The functions $N_q, q=1,2,3,4$ defined on the coordinate system $(s,t)$, and their corresponding first-order partial derivatives. \label{tbl:appd, N_i and derivatices}}
\end{table}

As described in Sec.~\ref{sec: spatial grid and discretization}, our simulation domain is $[a_x,b_x]\times[a_y,b_y]$, and our simulation grid is a rectangular grid with dimension $M\times N$, where each grid cell is a square with length $\Delta x=\Delta y = h$. Here, to keep our derivation general, we use $\Delta x$ and $\Delta y$ for now. For a grid cell indexed with $(i,j)$ in our simulation domain, as shown in Fig.~\ref{fig:FEM_appd_plot}(b), we can define functions on the grid cell region, $\phi_q(x,y)=N_q(s(x),t(y)), q=1,2,3,4$, where we have the transformation between $(s,t)$ and $(x,y)$,
\begin{equation}
\label{eqn:appd,fem,(x,y) and (s,t) transform}
x=a_x+\Delta x\left(i+\frac{s+1}{2}\right), \quad
y=a_y+\Delta y\left(j+\frac{t+1}{2}\right), \quad s,t\in[-1,1].
\end{equation}
Therefore, the functions $\phi_q(x,y)$ are linear on the grid cell region, have value $1$ at the corresponding corner node $q$, and have value $0$ at the other corner nodes. According to Eq.~\eqref{eqn:appd,fem,(x,y) and (s,t) transform}, we also have the derivatives
\begin{equation}
\label{eqn:appd,fem,(x,y) and (s,t) transform derivative}
\frac{dx}{ds}=\frac{\Delta x}{2}, \quad \frac{dy}{dt}=\frac{\Delta y}{2}, \quad \quad
\frac{ds}{dx}=\frac{2}{\Delta x}, \quad \frac{dt}{dy}=\frac{2}{\Delta y}.
\end{equation}
Define the cell region of grid cell $(i,j)$ to be $C_{i,j}$. Using Eq.~\eqref{eqn:appd,fem,(x,y) and (s,t) transform derivative}, for $p,q\in\{1,2,3,4\}$, we have the useful derivative calculations 
\begin{equation}
\label{eqn:appd,fem, phi derivative with (x,y)}
\begin{aligned}
    &\frac{\p \phi_q}{\p x}&&=\frac{\p N_q}{\p s}\cdot \frac{\p s}{\p x}=\frac{2}{\Delta x}\frac{\p N_q}{\p s}, \\
    &\frac{\p \phi_q}{\p y}&&=\frac{2}{\Delta y}\frac{\p N_q}{\p t}.
\end{aligned}
\end{equation}
From here we can derive the integral relations for a single $\phi_q$ as
\begin{equation}
  \label{eqn:appd,fem, phi derivative int with (x,y)}
  \begin{aligned}
    &\int_{C_{i,j}}\frac{\p \phi_q}{\p x}\,dx \, dy&&=\frac{\Delta y}{2}\int_{-1}^{1}\int_{-1}^{1}\frac{\p N_q}{\p s} \,ds \, dt,\\
    &\int_{C_{i,j}}\frac{\p \phi_q}{\p y}\,dx \, dy&&=\frac{\Delta x}{2}\int_{-1}^{1}\int_{-1}^{1}\frac{\p N_q}{\p t} \,ds \, dt
  \end{aligned}
\end{equation}
and products of $\phi_p$ and $\phi_q$ as
\begin{equation}
  \label{eqn:appd,fem, phi derivative int2 with (x,y)}
  \begin{aligned}
    &\int_{C_{i,j}}\frac{\p \phi_p}{\p x}\frac{\p \phi_q}{\p x}\,dx \, dy&&=\int_{-1}^{1}\int_{-1}^{1}\frac{2}{\Delta x}\frac{\p N_p}{\p s}\cdot \frac{2}{\Delta x}\frac{\p N_q}{\p s}\cdot \frac{\Delta x}{2}ds\frac{\Delta y}{2}dt \\
    &&&=\frac{\Delta y}{\Delta x}\int_{-1}^{1}\int_{-1}^{1}\frac{\p N_p}{\p s}\frac{\p N_q}{\p s} \,ds \, dt, \\
    &\int_{C_{i,j}}\frac{\p \phi_p}{\p y}\frac{\p \phi_q}{\p y}\,dx \, dy&&=\frac{\Delta x}{\Delta y}\int_{-1}^{1}\int_{-1}^{1}\frac{\p N_p}{\p t}\frac{\p N_q}{\p t} \,ds \, dt,\\
    &\int_{C_{i,j}}\phi_p\phi_q \,dx \, dy&&=\frac{\Delta x\Delta y}{4}\int_{-1}^{1}\int_{-1}^{1}N_p N_q \,ds \, dt, \\
    &\int_{C_{i,j}}\frac{\p \phi_p}{\p x}\frac{\p \phi_q}{\p y}\,dx \, dy&&=\int_{-1}^{1}\int_{-1}^{1}\frac{\p N_p}{\p s}\frac{\p N_q}{\p t} \,ds \, dt,\\
    &\int_{C_{i,j}}\frac{\p \phi_p}{\p y}\frac{\p \phi_q}{\p x}\,dx \, dy&&=\int_{-1}^{1}\int_{-1}^{1}\frac{\p N_p}{\p t}\frac{\p N_q}{\p s} \,ds \, dt.\\
\end{aligned}
\end{equation}

We can now introduce the nodal basis function, $\vpsi\in \mathbf{V}_h(\Omega)$, that is defined on the inner nodes and vanishes on the boundary $\Gamma$. $\vpsi=(\psi,0)$ for the $x$ dimension where $k=1$, and $\vpsi=(0,\psi)$ for the $y$ dimension where $k=2$. Here, $\psi\in V_h(\Omega)$ are chosen to be the scalar-valued bilinear element functions.

Suppose we are looking at an inner node in the simulation grid, $P_{\text{cc}}$. Its relationship with the adjacent nodes is shown in Fig.~\ref{fig:FEM_appd_plot}(d), where each adjacent node is indexed with two-letter suffix. The first letter is ``d'', ``c'', or ``u'', meaning down, center, and up, respectively. The second letter is ``l'', ``c'', or ``r'', meaning left, center, or right, respectively. The adjacent grid cell regions are labeled with $I, II, III, IV$. Then the scalar-valued bilinear element function $\psi_{\text{cc}}\in V_h$ for node $P_{\text{cc}}$ is composed of $\phi_1$ defined on grid cell region $I$, $\phi_2$ defined on $II$, $\phi_3$ defined on $III$, and $\phi_4$ defined on $IV$, as shown in Fig.~\ref{fig:FEM_appd_plot}(e). Therefore, $\psi_{\text{cc}}$ has value $1$ at node $P_{\text{cc}}$, is linear on the four adjacent regions of the node, and has value $0$ at all the neighboring nodes. Furthermore, the element $\psi_{\text{cc}}$ is only defined in the regions $I, II, III, IV$ surrounding the node $P_{\text{cc}}$, and is $0$ anywhere else in the simulation domain $\Omega$.

We now consider integral calculations of two elements. Since an element is only defined in the local regions of a node, the element can only have non-zero integral values with elements of its neighboring nodes and of itself, where they have overlapping regions. Let us call these nodes $P_{\delta}$. For example, for our node $P_{\text{cc}}$, its bilinear element $\psi_{\text{cc}}$ has overlapping region $IV$ with the bilinear element $\psi_{\text{ul}}$ of node $P_{\text{ul}}$. In region $IV$, $\psi_{\text{cc}}$ is given by $\phi_4$ and $\psi_{\text{ul}}$ is given by $\phi_1$. In Table~\ref{tbl:appd, overlapping elements composite functions in each region}, we summarize the overlapping regions for an element $\psi_{\text{cc}}$ with other elements $\psi_{\delta}$ in the domain, and their respective composition functions on the overlapping region, $\phi_{p}^{\text{cc}}$ for node $P_{\text{cc}}$ and $\phi_{q}^{\delta}$ for $P_{\delta}$. The table is helpful for fast integral evaluations involving two elements in the FEM derivation.

\begin{table}
\centering
\begin{tabular}{|cc|ccccccccc|}
\hline
\multicolumn{2}{|c|}{\multirow{2}{*}{$(p,q)$ for $(\phi_p^{\text{cc}},\phi_q^{\delta})$}}    & \multicolumn{9}{c|}{Nodes $P_{\delta}$ having overlapping elements with $P_{\text{cc}}$: neighbors and itself}    \\ \cline{3-11}
\multicolumn{2}{|c|}{}                     & \multicolumn{1}{c|}{$P_{\text{ul}}$} & \multicolumn{1}{c|}{$P_{\text{uc}}$} & \multicolumn{1}{c|}{$P_{\text{ur}}$} & \multicolumn{1}{c|}{$P_{\text{cl}}$} & \multicolumn{1}{c|}{$P_{\text{cc}}$} & \multicolumn{1}{c|}{$P_{\text{cr}}$} & \multicolumn{1}{c|}{$P_{\text{dl}}$} & \multicolumn{1}{c|}{$P_{\text{dc}}$} &  $P_{\text{dr}}$\\ \hline
\multicolumn{1}{|c|}{\multirow{4}{*}{Regions}} & $I$ & \multicolumn{1}{c|}{$(1,4)$} & \multicolumn{1}{c|}{$(1,3)$} & \multicolumn{1}{c|}{---} & \multicolumn{1}{c|}{$(1,2)$} & \multicolumn{1}{c|}{$(1,1)$} & \multicolumn{1}{c|}{---} & \multicolumn{1}{c|}{---} & \multicolumn{1}{c|}{---} & --- \\ \cline{2-11}
\multicolumn{1}{|c|}{}                  & $II$ & \multicolumn{1}{c|}{---} & \multicolumn{1}{c|}{$(2,4)$} & \multicolumn{1}{c|}{$(2,3)$} & \multicolumn{1}{c|}{---} & \multicolumn{1}{c|}{$(2,2)$} & \multicolumn{1}{c|}{$(2,1)$} & \multicolumn{1}{c|}{---} & \multicolumn{1}{c|}{---} & --- \\ \cline{2-11}
\multicolumn{1}{|c|}{}                  & $III$ & \multicolumn{1}{c|}{---} & \multicolumn{1}{c|}{---} & \multicolumn{1}{c|}{---} & \multicolumn{1}{c|}{$(3,4)$} & \multicolumn{1}{c|}{$(3,3)$} & \multicolumn{1}{c|}{---} & \multicolumn{1}{c|}{$(3,2)$} & \multicolumn{1}{c|}{$(3,1)$} &  --- \\ \cline{2-11}
\multicolumn{1}{|c|}{}                  & $IV$ & \multicolumn{1}{c|}{---} & \multicolumn{1}{c|}{---} & \multicolumn{1}{c|}{---} & \multicolumn{1}{c|}{---} & \multicolumn{1}{c|}{$(4,4)$} & \multicolumn{1}{c|}{$(4,3)$} & \multicolumn{1}{c|}{---} & \multicolumn{1}{c|}{$(4,2)$} & $(4,1)$ \\ \hline
\end{tabular}
\caption{For an element $\psi_{\text{cc}}$ defined on node $P_{\text{cc}}$, it only overlaps with itself, and with elements defined on its neighboring nodes. We collectively call these nodes $P_{\delta}$. The overlapping regions and the corresponding composition functions $\phi_q, q=1,2,3,4$ on the regions are given in this table. ``---'' means no overlap in the region.  \label{tbl:appd, overlapping elements composite functions in each region}}
\end{table}

As an example, suppose we want to evaluate the integral $\int_{\Omega}\frac{\p \psi_{\text{cc}}}{\p x}\frac{\p \psi_{\text{ul}}}{\p x}\,dx \, dy$, using Table~\ref{tbl:appd, overlapping elements composite functions in each region}, Eq.~\eqref{eqn:appd,fem, phi derivative int2 with (x,y)} and Table~\ref{tbl:appd, N_i and derivatices}, we have
\begin{equation}
\label{eqn:appd,fem,example integral evaluation of two neighboring elements}
\begin{aligned}
    \int_{\Omega}\frac{\p \psi_{\text{cc}}}{\p x}\frac{\p \psi_{\text{ul}}}{\p x}\,dx \, dy
    &= \int_{I}\frac{\p \phi_1}{\p x}\frac{\p \phi_4}{\p x}\,dx \, dy
    & \qquad \text{(using Table~\ref{tbl:appd, overlapping elements composite functions in each region}) } \\
    &= \frac{\Delta y}{\Delta x}\int_{-1}^{1} \int_{-1}^{1}\frac{\p N_1}{\p s}\frac{\p N_4}{\p s} \,ds \, dt
    & \qquad \text{(using Eq.~\eqref{eqn:appd,fem, phi derivative int2 with (x,y)}) } \\
    &= \frac{\Delta y}{\Delta x} \int_{-1}^{1} \int_{-1}^{1}\frac{(1-t)}{4}\frac{(-t-1)}{4} \,ds \, dt
    & \text{(using Table~\ref{tbl:appd, N_i and derivatices}) } \\
    &=-\frac{\Delta y}{6\Delta x}.
\end{aligned}
\end{equation}

\subsection{Components of $\vb$}
\label{appendix: fem: components of b}
$\vb$ is given by the source term on the right hand side of Eq.~\eqref{eqn:fem projection, main, rearrange}. Suppose we are looking at an inner node, $P_{\text{cc}}$, using the bilinear element $\psi_{\text{cc}}$ in each of the $x$ and $y$ dimensions. Then Eq.~\eqref{eqn:fem projection, main, rearrange} becomes
\begin{equation}
\label{eqn:appd,fem projection, main, rearrange, P_cc}
\begin{aligned}
\int_{\Omega}(\bC:\bD_{n+1}^I)_k\cdot\nabla\psi_{\text{cc}}(\vx)\,d\vx
&=
-\int_{\Omega}(\bC:\bD^B)_k\cdot\nabla\psi_{\text{cc}}(\vx)\,d\vx
-
\frac{1}{\Delta t}  \int_{\Omega}\bsig_{*,k}\cdot \nabla\psi_{\text{cc}} \,d\vx  , \\
\nabla\psi_{\text{cc}}&=
\begin{bmatrix}
    \p \psi_{\text{cc}}/ \p x \\ \p \psi_{\text{cc}}/ \p y
\end{bmatrix}, \quad k=1,2.
\end{aligned}
\end{equation}
Let the two integrals we need to evaluate on the right hand side be
\begin{equation}
    \label{eqn:appd,fem projection, b vector, A, B definition}
    A=\int_{\Omega}(\bC:\bD^B)_k\cdot\nabla\psi_{\text{cc}}(\vx)\,d\vx, \quad B=\int_{\Omega}\bsig_{*,k}\cdot \nabla\psi_{\text{cc}} \,d\vx.
\end{equation}
\begin{figure}
    \centering
    \includegraphics[width=1\textwidth]{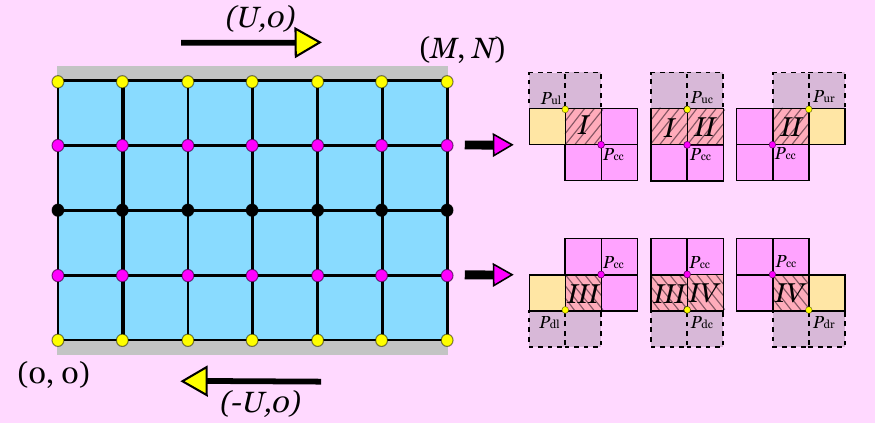}
    \caption{The overlapping regions between the element of an inner node $P_{\text{cc}}$ adjacent to the boundary, and the elements of its neighboring Dirichlet boundary nodes.}
    \label{fig:FEM_appd_bdry_inner_node_overlap}
\end{figure}%
First, let us consider the $x$ dimension, where $k=1$, and the nodal basis function for $P_{\text{cc}}$ in our simulation is $\vpsi_{\text{cc}}=(\psi_{\text{cc}},0)$. $A$ is non-zero only when $P_{\text{cc}}$ are nodes adjacent to the Dirichlet boundary nodes, and therefore the element $\psi_{\text{cc}}$ have overlapping region with bilinear elements defined on the Dirichlet boundary nodes. As shown in Fig.~\ref{fig:FEM_appd_bdry_inner_node_overlap}, the valid $P_{\text{cc}}$ can be adjacent to the Dirichlet boundary nodes on the upper and the lower walls of the simulation domain. Furthermore, for each valid $P_{\text{cc}}$, its bilinear element can overlap with the bilinear element of three Dirichlet boundary nodes. They are $P_{\text{ul}}$, $P_{\text{uc}}$, $P_{\text{ur}}$ for the upper wall, and $P_{\text{ll}}$, $P_{\text{dc}}$, $P_{\text{dr}}$ for the lower wall, as illustrated in Fig.~\ref{fig:FEM_appd_bdry_inner_node_overlap}.

Define $\alpha=K+\frac{4}{3}\mu$. The known Dirichlet boundary velocities are $\vv=(v_1^B,v_2^B)$. Since in simple shear, $v_2^B=0$ on both upper and lower walls, using Eq.~\eqref{eqn:fem projection, u and v n+1 expressed as basis functions}, $(\bC:\bD^B)_1$ given by Eq.~\eqref{eqn:CD tensor component} is simplified to be
\begin{equation}
    \label{eqn:appd,fem projection, CD_B, k=1}
    (\bC:\bD^B)_1=
    \begin{pmatrix}
    \alpha\sum_{j=1}^{\mathcal{B}}v_1^B\frac{\p \psi_j}{\p x} \quad \quad
    \mu\sum_{j=1}^{\mathcal{B}}v_1^B\frac{\p \psi_j}{\p y}
    \end{pmatrix}.
\end{equation}
Therefore, using Eq.~\eqref{eqn:appd,fem projection, CD_B, k=1}, $A$ in Eq.~\eqref{eqn:appd,fem projection, b vector, A, B definition} equals to
\begin{equation}
    \label{eqn:appd,fem projection, b vector, A calculation}
    \begin{aligned}
         A&=\int_{\Omega}\alpha\sum_{j=1}^{\mathcal{B}}v_1^B\frac{\p \psi_j}{\p x} \frac{\p \psi_{\text{cc}}}{\p x}\,dx \, dy + \int_{\Omega}\mu\sum_{j=1}^{\mathcal{B}}v_1^B\frac{\p \psi_j}{\p y} \frac{\p \psi_{\text{cc}}}{\p y} \,dx \, dy\\
         &=\alpha\sum_{j=1}^{\mathcal{B}}v_1^B\int_{\Omega}\frac{\p \psi_j}{\p x} \frac{\p \psi_{\text{cc}}}{\p x}\,dx \, dy + \mu\sum_{j=1}^{\mathcal{B}}v_1^B\int_{\Omega}\frac{\p \psi_j}{\p y} \frac{\p \psi_{\text{cc}}}{\p y} \,dx \, dy \\
         &=A_1+A_2.
    \end{aligned}
\end{equation}
When $P_{\text{cc}}$ are nodes adjacent to the upper/lower wall boundary nodes, the two terms of $A$ in Eq.~\eqref{eqn:appd,fem projection, b vector, A calculation} are calculated as
\begin{equation}
    \label{eqn:appd,fem projection, b vector, A calculation, upper wall, a,b}
         A_1= \alpha v_1^B \sum_{\delta} \left(
         \int_{\Omega} \frac{\p \psi_{\delta}}{\p x}\frac{\p \psi_{\text{cc}}}{\p x} \,dx \, dy \right),
         \qquad
         A_2= \mu v_1^B \sum_{\delta} \left(
         \int_{\Omega} \frac{\p \psi_{\delta}}{\p y}\frac{\p \psi_{\text{cc}}}{\p y} \,dx \, dy \right),
\end{equation}
where $\delta=\{\text{ul}, \text{uc}, \text{ur} \}$ when $P_{\text{cc}}$ are adjacent to the upper wall boundary nodes, and $\delta=\{\text{dl}, \text{dc}, \text{dr} \}$ when $P_{\text{cc}}$ are adjacent to the lower wall boundary nodes.

Equation~\eqref{eqn:appd,fem projection, b vector, A calculation, upper wall, a,b} can be evaluated by using the procedure same as the example integral calculation given in Eq.~\eqref{eqn:appd,fem,example integral evaluation of two neighboring elements}. For both upper and lower wall cases, we have
\begin{equation}
    \label{eqn:appd,fem projection, b vector, A calculation, both walls, end eval}
    A=A_1+A_2=\alpha v_1^B\cdot 0 -\mu v_1^B \frac{\Delta x}{\Delta y}=-\frac{\Delta x}{\Delta y}\mu v_1^B.
\end{equation}
Now, let us look at the calculation of $B$ defined in Eq.~\eqref{eqn:appd,fem projection, b vector, A, B definition} in the $x$ dimension, $k=1$. $B$ is non-zero for any inner node $P_{\text{cc}}$. From Eq.~\eqref{eqn:stress tensor component}, we know $\bsig_{*,1}$ is given by
\begin{equation}
    \label{eqn:appd,fem projection, stress_*, k=1}
    \bsig_{*,1}=
    \begin{pmatrix}
    -p_*+s_*-q_* \quad \quad
    \tau_*
    \end{pmatrix}
\end{equation}
and therefore, we have
\begin{equation}
    \label{eqn:appd,fem projection, B calculation}
    \begin{aligned}
         B&=\int_{\Omega}(-p_*+s_*-q_*)\frac{\p \psi_{\text{cc}}}{\p x}\,dx \, dy
            +\int_{\Omega}\tau_*\frac{\p \psi_{\text{cc}}}{\p y}\,dx \, dy \\
          &=\int_{I}(-p_*+s_*-q_*)\frac{\p \phi_1}{\p x}+\tau_*\frac{\p \phi_1}{\p y}\,dx \, dy
            +\int_{II}(-p_*+s_*-q_*)\frac{\p \phi_2}{\p x}+\tau_*\frac{\p \phi_2}{\p y}\,dx \, dy \\
            &+\int_{III}(-p_*+s_*-q_*)\frac{\p \phi_3}{\p x}+\tau_*\frac{\p \phi_3}{\p y}\,dx \, dy
            +\int_{IV}(-p_*+s_*-q_*)\frac{\p \phi_4}{\p x}+\tau_*\frac{\p \phi_4}{\p y}\,dx \, dy \\
          &=\frac{\Delta y}{2}(-p_*^{I}+s_*^{I}-q_*^{I})-\frac{\Delta x}{2}\tau_*^{I}
          -\frac{\Delta y}{2}(-p_*^{II}+s_*^{II}-q_*^{II})-\frac{\Delta x}{2}\tau_*^{II} \\
          &+\frac{\Delta y}{2}(-p_*^{III}+s_*^{III}-q_*^{III})+\frac{\Delta x}{2}\tau_*^{III}
          -\frac{\Delta y}{2}(-p_*^{IV}+s_*^{IV}-q_*^{IV})+\frac{\Delta x}{2}\tau_*^{IV},
    \end{aligned}
\end{equation}
where $I, II, III, IV$ are the surrounding grid cell regions of $P_{\text{cc}}$, as shown in Fig.~\ref{fig:FEM_appd_plot}(d). The superscript on the stress components above denotes the stress in the corresponding region.

Now that we have calculated $A$ and $B$ for $k=1$ defined in Eq.~\eqref{eqn:appd,fem projection, b vector, A, B definition}, we can calculate the components of $\vb$ given by the right hand side of Eq.~\eqref{eqn:appd,fem projection, main, rearrange, P_cc} for the $x$-component elements $\vpsi_{\text{cc}}=(\psi_{\text{cc}},0)$.

For the $y$ dimension, where $k=2$, and the nodal basis function for $P_{\text{cc}}$ in our simulation is $\vpsi_{\text{cc}}=(0,\psi_{\text{cc}})$. Following the same procedure as above, we obtain, for all inner nodes,
\begin{equation}
    \label{eqn:appd,fem projection, k=2, A,B results}
    \begin{aligned}
         A&=0,\\
         B&=\frac{\Delta y}{2}\tau_*^{I}-\frac{\Delta x}{2}(-p_*^{I}-s_*^{I}-q_*^{I})
          -\frac{\Delta y}{2}\tau_*^{II}-\frac{\Delta x}{2}(-p_*^{II}-s_*^{II}-q_*^{II}) \\
          &+\frac{\Delta y}{2}\tau_*^{III}+\frac{\Delta x}{2}(-p_*^{III}-s_*^{III}-q_*^{III})
          -\frac{\Delta y}{2}\tau_*^{IV}+\frac{\Delta x}{2}(-p_*^{IV}-s_*^{IV}-q_*^{IV}).
    \end{aligned}
\end{equation}
Therefore, we can compute the entries of $\vb$ for the $y$-component elements, $\vpsi_{\text{cc}}=(0,\psi_{\text{cc}})$.

\subsection{Components of $\bA$ and $\vw$}
\label{appendix: fem: components of Aw}

$\bA\vw$ is given by the left hand side of Eq.~\eqref{eqn:fem projection, main, rearrange}. Let $k=1,2$ represent the $x$ and $y$ dimensions, and $i=1,\ldots,\mathcal{I}$ the inner nodes. The inner node velocity weights $w_{i,k}$ make up the entries of $\vw$, and $\vw=[w_{1,1} \quad w_{1,2}\quad w_{2,1}\quad w_{2,2}\quad  \ldots\quad w_{\mathcal{I},1}\quad w_{\mathcal{I},2}]^\trans$.
Suppose we are looking at an inner node $P_{\text{cc}}$. Its bilinear element function in each dimension is $\psi_{\text{cc}}$. We can calculate its corresponding row entry in $\bA$ using the left hand side of Eq.~\eqref{eqn:appd,fem projection, main, rearrange, P_cc}.
First, when $k=1$, and our nodal basis function of the simulation is $\vpsi_\text{cc}=(\psi_{\text{cc}},0)$. Define $\beta=K-\frac{2}{3}\mu$.
Then, using Eq.~\eqref{eqn:fem projection, u and v n+1 expressed as basis functions}, $(\bC:\bD_{n+1}^I)_1$ given by Eq.~\eqref{eqn:CD tensor component} is
\begin{equation}
    \label{eqn:appd,fem projection, CD_{n=1}, k=1, components}
    (\bC:\bD_{n+1}^I)_1=
    \begin{pmatrix}
    \alpha \sum_{i=1}^{\mathcal{I}}w_{i,1}\frac{\p \psi_i}{\p x}+ \beta \sum_{j=1}^{\mathcal{I}}w_{j,2}\frac{\p \psi_j}{\p y}
     \quad \quad
    \mu\left( \sum_{j=1}^{\mathcal{I}}w_{j,2}\frac{\p \psi_j}{\p x} + \sum_{i=1}^{\mathcal{I}}w_{i,1}\frac{\p \psi_i}{\p y} \right)
    \end{pmatrix}.
\end{equation}
Therefore the left hand side of Eq.~\eqref{eqn:appd,fem projection, main, rearrange, P_cc} is
\begin{equation}
    \label{eqn:appd,fem projection, matrix A, calculation expansion}
    \begin{aligned}
    \int_{\Omega}(\bC:\bD_{n+1}^I)_1\cdot \nabla \psi_{\text{cc}}(\vx)\,dx \, dy
    &= \int_{\Omega} (\alpha \sum_{i=1}^{\mathcal{I}}w_{i,1}\frac{\p \psi_i}{\p x})\frac{\p \psi_{\text{cc}}}{\p x} \,dx \, dy + \int_{\Omega} (\beta \sum_{j=1}^{\mathcal{I}}w_{j,2}\frac{\p \psi_j}{\p y})\frac{\p \psi_{\text{cc}}}{\p x} \,dx \, dy \\
    &+\int_{\Omega} (\mu \sum_{j=1}^{\mathcal{I}}w_{j,2}\frac{\p \psi_j}{\p x})\frac{\p \psi_{\text{cc}}}{\p y} \,dx \, dy + \int_{\Omega} (\mu \sum_{i=1}^{\mathcal{I}}w_{i,1}\frac{\p \psi_i}{\p y})\frac{\p \psi_{\text{cc}}}{\p y} \,dx \, dy \\
    &=A+B+C+D.
    \end{aligned}
\end{equation}
The terms $A,B,C,D$ in Eq.~\eqref{eqn:appd,fem projection, matrix A, calculation expansion} can be evaluated separately and then put together. The element $\psi_{\text{cc}}$ only has non-zero integral values with itself and its neighboring inner node elements, as given in Table~\ref{tbl:appd, overlapping elements composite functions in each region}. We give an example evaluation of the first integral term,
\begin{equation}
    \label{eqn:appd,fem projection, matrix A, calculation expansion, A calculation}
    A=\alpha \sum_{i=1}^{\mathcal{I}}w_{i,1} \int_{\Omega} \frac{\p \psi_i}{\p x}\frac{\p \psi_{\text{cc}}}{\p x}dx\,dy
    = \alpha \sum_{\delta} \left(
    w_{\delta,1}\int_{\Omega} \frac{\p \psi_{\delta}}{\p x}  \frac{\p \psi_{\text{cc}}}{\p x} dx\,dy
      \right),
\end{equation}
where the sum is taken over nine values, $\delta=\bigl\{\text{ul}, \text{uc}, \text{ur},\text{cl},\text{cc},\text{cr},\text{dl},\text{dc},\text{dr} \bigl\}$.
Putting $A,B,C,D$ together, Eq.~\eqref{eqn:appd,fem projection, matrix A, calculation expansion} becomes
\begin{equation}
    \label{eqn:appd,fem projection, matrix A, calculation expansion, dimension 1, A+B+C+D result}
    \begin{aligned}
    \int_{\Omega}(\bC:\bD_{n+1}^I)_1\cdot \nabla \psi_{\text{cc}}(\vx)\,d\vx
    &=\sum_{\delta} \left[ w_{\delta,1}\left( \alpha\int_{\Omega}\frac{\p \psi_{\delta}}{\p x}\frac{\p \psi_{\text{cc}}}{\p x}\,d\vx +\mu \int_{\Omega}\frac{\p \psi_{\delta}}{\p y} \frac{\p \psi_{\text{cc}}}{\p y}\,d\vx \right) \right.
    \\
    &\left. \quad\quad\quad + w_{\delta,2} \left( \beta\int_{\Omega}\frac{\p \psi_{\delta}}{\p y}\frac{\p \psi_{\text{cc}}}{\p x}\,d\vx +\mu \int_{\Omega}\frac{\p \psi_{\delta}}{\p x} \frac{\p \psi_{\text{cc}}}{\p y}\,d\vx \right) \right].
    \end{aligned}
\end{equation}
The integrals in Eq.~\eqref{eqn:appd,fem projection, matrix A, calculation expansion, dimension 1, A+B+C+D result} can be evaluated following the example integral calculation procedure given in Eq.~\eqref{eqn:appd,fem,example integral evaluation of two neighboring elements}. Then we can obtain the corresponding row entries in $\bA$ for the element $\vpsi_{\text{cc}}=(\psi_{\text{cc}}, 0)$ defined on inner node $P_{\text{cc}}$, with corresponding unknown weight $w_{\text{cc},1}$. The row is mostly $0$, and only non-zero for entries corresponding to the elements of $P_{\delta}$, as given in Table~\ref{tbl:appd, overlapping elements composite functions in each region}.

When $k=2$, the nodal basis function for inner node $P_{\text{cc}}$ is $\vpsi_{\text{cc}}=(0, \psi_{\text{cc}})$, with corresponding unknown weight $w_{\text{cc},2}$. We can follow the same procedure as above to obtain the corresponding row entries.

We can use a simple format to represent the entries for the sparse matrix $\bA$. For an inner node $P_{\text{cc}}$, let a $2\times2$ matrix $a_{\delta}$ represent the non-zero coefficients in $\bA$ between the nodal basis elements $\vpsi_{\text{cc}}$ defined on $P_{\text{cc}}$, and the elements $\vpsi_{\delta}$ that have overlapping regions with $\vpsi_{\text{cc}}$. These elements $\vpsi_{\delta}$ are defined on nodes $P_{\delta}$ given in Table~\ref{tbl:appd, overlapping elements composite functions in each region}. Let the first row represent the coefficients for $k=1$ and $\vpsi_{\text{cc}}=(\psi_{\text{cc}},0)$; let the second row be the coefficients for $k=2$ and $\vpsi_{\text{cc}}=(0, \psi_{\text{cc}})$. Let the first column be coefficients of $\vpsi_{\text{cc}}$ with the overlapping element function in the $x$-dimension, $\vpsi_{\delta}=(\psi_{\delta},0)$; let the second column be the coefficients of $\vpsi_{\text{cc}}$ with $\vpsi_{\delta}=(0,\psi_{\delta})$. Then matrix $\bA$ has components given by
\begin{equation}
\label{eqn:appd,fem projection, matrix A, simple format component forms}
\begin{aligned}
a_{\text{ul}}=a_{\text{dr}}&=
\begin{pmatrix}
   -\frac{\mu \Delta x}{6 \Delta y}-\frac{\alpha\Delta y}{6\Delta x}  & \frac{K}{4}+\frac{\mu}{12} \\
   \frac{K}{4}+\frac{\mu}{12} & -\frac{\alpha \Delta x}{6 \Delta y}-\frac{\mu \Delta y}{6 \Delta x}
\end{pmatrix}, \quad
&&a_{\text{uc}}=a_{\text{dc}}=
\begin{pmatrix}
   -\frac{2\mu\Delta x}{3\Delta y}+\frac{\alpha \Delta y}{3\Delta x}  & 0 \\
   0 & -\frac{2\alpha \Delta x}{3\Delta y}+\frac{\mu \Delta y}{3 \Delta x}
\end{pmatrix}, \\
a_{\text{ur}}=a_{\text{dl}}&=
\begin{pmatrix}
   -\frac{\mu \Delta x}{6 \Delta y}-\frac{\alpha\Delta y}{6\Delta x}  & -\frac{K}{4}-\frac{\mu}{12} \\
   -\frac{K}{4}-\frac{\mu}{12} & -\frac{\alpha \Delta x}{6 \Delta y}-\frac{\mu \Delta y}{6 \Delta x}
\end{pmatrix}, \quad
&&a_{\text{cl}}=a_{\text{cr}}=
\begin{pmatrix}
   \frac{\mu\Delta x}{3\Delta y}-\frac{2\alpha \Delta y}{3\Delta x}  & 0 \\
   0 & \frac{\alpha \Delta x}{3\Delta y}-\frac{2\mu \Delta y}{3 \Delta x}
\end{pmatrix}, \\
a_{\text{cc}}&=
\begin{pmatrix}
   \frac{4\mu\Delta x}{3\Delta y}+\frac{4\alpha\Delta y}{3\Delta x}  & 0 \\
   0 & \frac{4\alpha \Delta x}{3 \Delta y}+\frac{4\mu \Delta y}{3 \Delta x}
\end{pmatrix}.
\end{aligned}
\end{equation}
The components in Eq.~\eqref{eqn:appd,fem projection, matrix A, simple format component forms} are further simplified in our simulation since $\Delta x =\Delta y = h$.

\section{FEM projection linear system derivation: Second-order projection method}
\label{appendix: FEM second-order projection}
The FEM projection for the second-order projection method described in Sec.~\ref{sec:Second-order projection method for quasi-static elastoplasticity} follows similar derivation to the FEM projection for the original projection method. The only difference is that we use boundary velocities, $\bPhi_B=\mathbf{0}$, as Dirichlet boundary condition, for solving the correction velocity term, $\bPhi_*$. This simplifies the calculation of the $\vb$ term, as in Eq.~\eqref{eqn:appd,fem projection, b vector, A, B definition}, $A=0$ for both $x$ and $y$ dimensions, corresponding to $k=1,2$. We therefore just need to evaluate the $B$ term.

\bibliographys{refs}

\end{document}